\documentclass[conference,9pt]{IEEEtran}
\IEEEoverridecommandlockouts
\usepackage{amsmath,amssymb,amsfonts}

\usepackage{graphicx}
\usepackage{textcomp}
\usepackage{xcolor}
\usepackage{blindtext}
\usepackage{geometry}
\usepackage{authblk}
\usepackage{tabularx}
\usepackage{lipsum}

\usepackage{textcomp}
\usepackage{amsthm}
\usepackage{svg}
\usepackage{booktabs}
\usepackage{multirow}
\usepackage{multicol}
\usepackage[colorlinks=true, citecolor= blue]{hyperref}
\usepackage[numbers]{natbib}
 \usepackage{setspace}
\usepackage[all]{xy}
\usepackage{fancyhdr}
\usepackage[active]{srcltx}
\usepackage{mathptmx}

\fancypagestyle{noheader}{}
\usepackage{hyperref} 
\usepackage{natbib} 
\usepackage{listings} 
\usepackage{algorithm} 
\usepackage{algpseudocode} 
\usepackage{xcolor} 
\usepackage[toc]{glossaries} 
\usepackage{xspace}
\usepackage{helvet}
 
\emergencystretch=\maxdimen
\definecolor{codegray}{rgb}{0.95,0.95,0.95}  
\definecolor{codepurple}{rgb}{0.58, 0, 0.82}   
\definecolor{codegreen}{rgb}{0,0.6,0}   
\definecolor{codenumbers}{rgb}{0.5,0.5,0.5}  

\lstdefinestyle{mystyle}{
    backgroundcolor=\color{codegray},   
    commentstyle=\color{codegreen},
    keywordstyle=\color{magenta},
    numberstyle=\color{codenumbers},
    stringstyle=\color{codepurple},
    basicstyle=\ttfamily\footnotesize,
    breakatwhitespace=false,         
    breaklines=true,                 
    captionpos=b,                    
    keepspaces=true,                 
    numbers=left,                    
    numbersep=5pt,                  
    showspaces=false,                
    showstringspaces=false,
    showtabs=false,                  
    tabsize=2
}
\title{Parametric Amplification in Kerr Nonlinear Resonators: A theoretical review of Josephson Parametric Amplifiers}

\author[1*]{Rajlaxmi Bhoite\thanks{$^1$rajlaxmivilasbhoite@iisertirupati.ac.in}}
\author[2]{Shraddhanjali Choudhury\thanks{$^2$ph22c046@smail.iitm.ac.in}}

\affil[1]{Indian Institute of Science Education and Research (IISER), Tirupati, India}
\affil[2]{Indian Institute of Technology Madras, Chennai, India}

\date{28 July 2025}

\begin{document}

\maketitle
\begin{abstract}
This paper presents a detailed theoretical review of the amplification process in Josephson Parametric Amplifiers (JPAs), which are crucial for quantum-limited signal amplification in superconducting circuits \cite{Yurke2005,Esposito2019}. The paper begins by outlining the principles of parametric amplification, focusing on how a strong classical pump interacts with the nonlinear Josephson medium in reflection geometry. The key dynamical equations are derived under intense pumping, leading to a nonlinear steady-state solution \cite{Yurke2005}. Linearization around this solution allows to analyze the system response to weak signals and extract expressions for parametric gain and intermodulation gain using the input-output formalism \cite{Eichler2014,Castellanos2007}. Numerically, the equations are solved to explore how gain depends on frequency detuning and pump strength, which is visualized with a gain response curve. By enhancing our understanding of JPAs, this work aims to inspire continued research in the field of quantum-limited amplification.
\end{abstract}
\section{Motivation} 
Josephson Parametric Amplifiers (JPAs) are essential for amplifying weak microwave signals with near quantum limited noise performance an ability critical for high-fidelity readout in quantum information processing. Unlike conventional amplifiers such as HEMTs, JPAs offer high gain (often $>$20dB) \cite{Yurke2005} with tunable frequency and gain, enabling selective signal amplification with minimal added noise. JPAs are crucial in superconducting qubit readout, allowing single-shot measurements, and they also enable the generation and detection of squeezed microwave states, important for quantum metrology \cite{Jeffrey2014,Walter2017}. Their non-dissipative Josephson nonlinearity underpins efficient parametric amplification, setting them apart from other devices with lossy nonlinearities. Additionally, JPAs are compact, on-chip components, easily integrated into superconducting circuits—an advantage for scalable quantum computing. Beyond applications, they offer insight into nonlinear quantum dynamics, bistability, and quantum noise squeezing, making them valuable both technologically and fundamentally.
\section{What is a JPA}
The Josephson Parametric Amplifier is a superconducting quantum-limited amplifier that uses the nonlinear inductance of Josephson junctions to amplify microwave signals. These amplifiers are critical components in quantum computing and measurement systems, particularly for reading out qubits with minimal added noise \cite{Jeffrey2014}. A quantum-limited amplifier adds the minimum possible noise allowed by quantum mechanics to a signal during amplification. This is a fundamental limit, not due to imperfections, but imposed by the Heisenberg uncertainty principle. This is achieved when the amplifier is impedance-matched and internal losses are negligible. For phase-preserving JPAs, the minimum noise added is $N_{\text{add}} = \frac{1}{2}$. The term 'parametric' refers to a system in which some parameter (such as the refractive index or impedance) is being varied over time, often by an external pump field. In both optical and microwave circuits, this external modulation leads to energy being transferred between different frequency modes, typically creating or amplifying photons\\
\\
In a non-linear optical medium, the polarization \( \mathbf{P} \) (response of the material to an electric field \( \mathbf{E} \)) is not proportional to the field. It includes higher-order terms in the electric field:

\begin{align}
\mathbf{P} = \varepsilon_0 \left( \chi^{(1)} \mathbf{E} + \chi^{(2)} \mathbf{E}^2 + \chi^{(3)} \mathbf{E}^3 + \dots \right)
\end{align}

\begin{itemize}
    \item \( \chi^{(2)} \): leads to three-wave mixing
    \item \( \chi^{(3)} \): leads to four-wave mixing
\end{itemize}
While the second-order nonlinear susceptibility $\chi^{(2)}$ is nonzero only in materials that lack inversion symmetry, Josephson junctions can emulate an effective $\chi^{(2)}$ like interaction through flux modulation.  
\noindent
In Three-Wave Mixing (\( \chi^{(2)} \)) one pump photon (frequency \( \omega_p \)) is split into the following:
\begin{itemize}
    \item One signal photon (\( \omega_s \))
    \item One idler photon (\( \omega_i \))
\end{itemize}
\noindent
The pump photon is a photon from a strong, external electromagnetic field (often a microwave or optical tone). Its role is to pump energy into a non-linear system (like a Josephson junction or SQUID), enabling frequency conversion or amplification of weaker signals. It is not the signal that is being amplified— it is the energy source that drives the nonlinear interaction. The pump photon comes from an external generator or oscillator. For three wave mixing:
\begin{equation}
\omega_p = \omega_s + \omega_i
\label{eq:3wm_energy}
\end{equation}

\noindent
This process is called spontaneous parametric down-conversion (SPDC) when the signal and idler fields start in the vacuum state. \\
\\
\noindent
In Four-Wave Mixing (\( \chi^{(3)} \)) two pump photons combine to produce:
\begin{itemize}
    \item One signal photon (\( \omega_s \))
    \item One idler photon (\( \omega_i \))
\end{itemize}
\noindent
This is called Spontaneous four-wave mixing (SFWM) when the signal and idler fields start in a vacuum. We have:
\begin{equation}
2\omega_p = \omega_s + \omega_i
\label{eq:4wm_energy}
\end{equation}
\noindent
In optical systems, nonlinear crystals (with nonlinearity) allow to mix frequencies of light, creating signal and idler photons via processes like SPDC or SFWM. In the microwave regime, one wants to do the same — mix or amplify microwave-frequency signals — but there are no natural nonlinear crystals. Instead, engineered quantum circuits are used, particularly Josephson junctions and SQUIDs, to achieve nonlinearity. These artificial atoms play the role of non-linear media in cQED. Photons in the GHz frequency range are quantized excitations of the electromagnetic field, much like visible light photons, but at much lower frequencies. Although they are not light in the visible sense, they are still electromagnetic in nature. In this context, the concept of refractive index used in optics is replaced by circuit parameters like inductance and capacitance, which together determine the impedance of superconducting circuits that generate and manipulate these microwave photons. To realize a parametric process in circuits, the impedance needs to be modulated, typically via the inductance, using a Josephson junction. A Josephson junction has a Josephson inductance given approximately by:
\[
L_J(t) \approx \frac{\Phi_0}{2\pi I_c \cos\left({\phi(t)} \right)} \quad
\]

where:
\begin{itemize}
    \item $\Phi_0 = \frac{h}{2e}$ is the flux quantum,
    \item $I_c$ is the critical current,
    \item $\phi(t)$ is the superconducting phase difference across the junction.
\end{itemize}
This inductance can be derived from the Josephson relations: \( I = I_c \sin \phi \) and \( V = \frac{\hbar}{2e} \frac{d\phi}{dt} = \frac{\Phi_0}{2\pi} \frac{d\phi(t)}{dt}
    \). The Josephson inductance \( L_J(t) \) is defined by the small-signal linear relation:
\[
V(t) = L_J(t) \frac{dI(t)}{dt}
\]
\noindent
Using the current-phase relation:
\[
I(t) = I_c \sin \phi(t) \Rightarrow \frac{dI}{dt} = I_c \cos \phi(t) \cdot \frac{d\phi}{dt}
\]
\noindent
From the voltage relation:
\[
\frac{d\phi}{dt} = \frac{2\pi}{\Phi_0} V(t)
\]
\noindent
Substituting into the derivative:
\[
\frac{dI}{dt} = I_c \cos \phi(t) \cdot \frac{2\pi}{\Phi_0} V(t)
\]
\noindent
Finally,
\[
V(t) = \frac{1}{I_c \cos \phi(t)} \cdot \frac{\Phi_0}{2\pi} \cdot \frac{dI}{dt}
\]
\noindent
Comparing with \( V(t) = L_J(t) \frac{dI}{dt} \), we identify:
\begin{equation}
L_J(t) = \frac{\Phi_0}{2\pi I_c \cos \phi(t)}
\label{eq:LJ}
\end{equation}
\noindent
In the small current limit $I(t) \ll I_c$, the inductance can be determined as follows. \\
\\
\noindent
The cosine of the phase is:
\[
\cos \phi(t) = \sqrt{1 - \left( \frac{I(t)}{I_c} \right)^2}
\]
\noindent
Substituting into the inductance expression (see Eq.~\eqref{eq:LJ})
\[
L_J(t) = \frac{\Phi_0}{2\pi I_c \sqrt{1 - \left( \frac{I(t)}{I_c} \right)^2}}
\] 


\noindent 
Expanding the denominator using the binomial expansion:
\[
(1 - x^2)^{-1/2} \approx 1 + \frac{1}{2}x^2 + \frac{3}{8}x^4 + \cdots
\]
\noindent 
Keeping only the leading nonlinear term:
\begin{equation}
L_J(t) \approx L_J \left( 1 + \frac{1}{2} \left( \frac{I(t)}{I_c} \right)^2 \right)
\label{eq:LJ_nonlinear}
\end{equation}

\noindent 
where the zero-current (linear) inductance:
\[
L_J = \frac{\Phi_0}{2\pi I_c}
\]
\noindent 
This shows that the inductance depends \textit{nonlinearly} on the current. If the junction is driven with a microwave signal (i.e., an AC current), the inductance becomes time-varying. This implies a time-varying impedance — analogous to a modulated refractive index in optics. This periodic modulation enables four-wave mixing (FWM), in which:

\begin{equation} \text{Pump} + \text{Pump} \rightarrow \text{Signal} + \text{Idler} \label{eq:FWM} \notag \end{equation}

\noindent
In a SQUID, the effective Josephson inductance can be tuned by applying an external magnetic flux $\Phi_{\text{ext}}(t)$ through the loop:

\begin{equation}
L_{\text{SQUID}}(t) \approx L_J \left(1 + \frac{I(t)}{I_0} \right)
\label{eq:SQUID_L}
\end{equation}

where:
\begin{itemize}
    \item $I(t)$ is the AC drive current,
    \item $I_0$ depends on the DC flux bias applied to the SQUID loop.
\end{itemize}
\noindent
Since the modulation is linear in $I(t)$ (not quadratic), it can support three-wave mixing:

\[
\text{Pump} \rightarrow \text{Signal} + \text{Idler}
\]
The time-dependent, nonlinear inductance can be used to construct parametric amplifiers, which amplify weak microwave signals by mixing them with a strong pump field. The following description corresponds to four-wave mixing, with the model from \cite{Yurke2005} and the derivations demonstrated in detail, further supported by a computational study demonstrating 20dB of parametric gain for a JPA.
\section{The Hamiltonian}
\subsection*{1. Hamiltonian of the Nonlinear Resonator}
The Hamiltonian of a nonlinear resonator, with Kerr nonlinearity $K$, is given by:
\begin{equation}
H_r = \hbar \omega_0 A^\dagger A + \frac{\hbar}{2} K A^\dagger A^\dagger AA
\label{eq:H_resonator}
\end{equation}
Here:
\begin{itemize}
    \item $\omega_0$ is the resonant frequency of the cavity.
    \item $A$ and $A^\dagger$ are the annihilation and creation operators of the cavity mode.
    \item $K$ is the Kerr constant, representing the strength of the nonlinearity.
\end{itemize}
\begin{figure}[H]
    \centering
    \includegraphics[width=1\linewidth]{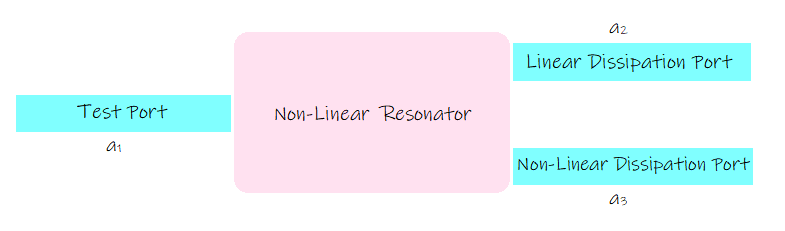}
    \caption{The model consists of an input/output port(bath) a$_1$ coupled ($\gamma_1$) to a non-linear resonator. The loss/dissipation is modeled via port a$_2$ and port a$_3$ (two photon loss) with coupling $\gamma_2$ and $\gamma_3$ to the resonator.}
    \label{fig:enter-label}
\end{figure}
\noindent
This Hamiltonian describes a system where the energy levels are not equally spaced due to the Kerr effect, leading to phenomena such as photon blockade and bistability. In this configuration, the resonator operates as an amplifier, where the reflected signal from the input port \( a_1 \) is stronger than the incoming signal. This is known as the \textit{negative-resistance reflection mode} at microwave frequencies. To model energy dissipation theoretically, two additional fictitious ports are introduced:

\begin{itemize}
  \item Port \( a_2 \): represents linear (single-photon) dissipation.
  \item Port \( a_3 \): accounts for nonlinear (two-photon) dissipation and is nonlinearly coupled to the resonator mode \( A \).
\end{itemize}

\subsection*{2. Total System Hamiltonian}
\noindent
The total Hamiltonian includes the resonator $H_r$ and its interaction with various ports:
\begin{equation}
H = H_r + H_{a1} + H_{a2} + H_{a3} + H_{T1} + H_{T2} + H_{T3}
\label{total}
\end{equation}
Where:
\begin{itemize}
    \item $H_{a1}, H_{a2}, H_{a3}$ are the Hamiltonians of the external baths (ports).
    \item $H_{T1}, H_{T2}, H_{T3}$ are the interaction Hamiltonians between the resonator and the respective ports.
\end{itemize}
\noindent
Each bath is modeled as a continuum of harmonic oscillators:
\begin{align}
H_{a1} &= \int d\omega \, \hbar \omega \, a_1^\dagger(\omega) a_1(\omega) \\
H_{a2} &= \int d\omega \, \hbar \omega \, a_2^\dagger(\omega) a_2(\omega) \\
H_{a3} &= \int d\omega \, \hbar \omega \, a_3^\dagger(\omega) a_3(\omega)
\end{align}
\noindent
Here, $a_j(\omega)$ and $a_j^\dagger(\omega)$ are the annihilation and creation operators for the bath modes at frequency $\omega$. The interactions between the resonator and the baths are given by:
\noindent
\subsubsection*{Linear Coupling (Ports a1 and a2)}
\begin{align}
H_{T1} &= \hbar \int d\omega \left[ \kappa_1 A^\dagger a_1(\omega) + \kappa_1^* a_1^\dagger(\omega) A \right] \\
H_{T2} &= \hbar \int d\omega \left[ \kappa_2 A^\dagger a_2(\omega) + \kappa_2^* a_2^\dagger(\omega) A \right]
\end{align}
\noindent
\subsubsection*{Nonlinear (Two-Photon) Coupling (Port a3)}
\begin{equation}
H_{T3} = \hbar \int d\omega \left[ \kappa_3 A^\dagger A^\dagger a_3(\omega) + \kappa_3^* a_3^\dagger(\omega) A A \right]
\end{equation}
\noindent
Here, the magnitude \( |\kappa_i| \) determines the coupling strength (rate of energy exchange), while the phase of \( \kappa_i \) encodes interference or coherence effects and can be tunable via external parameters such as drive or flux. The parameters \( \kappa_1 \) and \( \kappa_2 \) correspond to the coupling constant for linear couplings between the system and external ports, while \( \kappa_3 \) is the nonlinear (two-photon) coupling constant. The two-photon coupling represents processes where two photons in the resonator are annihilated simultaneously, creating a single photon in the bath.
\section*{Heisenberg Equation of Motion for Annihilation Operator $A$}
\noindent
The Heisenberg equation of motion is:
\begin{equation}
\frac{dA}{dt} = \frac{1}{i\hbar} [A, H]
\label{eq:EOM_A}
\end{equation}
\noindent
and the total Hamiltonian (Eq.~\eqref{total}):
\begin{equation}
H = H_r + H_{a1} + H_{a2} + H_{a3} + H_{T1} + H_{T2} + H_{T3}
\notag
\end{equation}
\noindent
The resonator Hamiltonian is given by $H_r$ (Eq.~\eqref{eq:H_resonator})

\begin{equation}
H_r = \hbar\omega_0 A^\dagger A + \frac{\hbar}{2} K A^\dagger A^\dagger A A \notag
\end{equation}
\noindent
\textit{(a) First term:}
\begin{align}
[A, \hbar\omega_0 A^\dagger A] &= \hbar\omega_0 [A, A^\dagger A] \notag \\ 
&= \hbar\omega_0 ([A, A^\dagger] A + A^\dagger [A, A]) = \hbar\omega_0 A
\end{align}
\noindent
\textit{(b) Second term:}
\begin{align}
[A, \frac{\hbar}{2} K A^\dagger A^\dagger A A] &= \frac{\hbar}{2} K [A, A^\dagger A^\dagger A A] \notag \\ 
&= \frac{\hbar}{2} K \cdot 2 A^\dagger A A = \hbar K A^\dagger A A
\end{align}

So:
\begin{equation}
[A, H_r] = \hbar\omega_0 A + \hbar K A^\dagger A A
\end{equation}

\subsection*{Commutator with Bath Hamiltonians $H_{ai}$}

\begin{equation}
[A, H_{ai}] = 0 \quad
\end{equation}
These are integrals over bath-only operators, which commute with A since they're different systems. So these terms do not contribute. Because the bath degrees of freedom act on a different Hilbert space than the system operator A. They're independent systems, so their operators commute.

\subsection*{Commutators with Interaction Hamiltonians}
\noindent
There are three types of Hamiltonians that couple a system operator $A$ to bath modes $a_i(\omega)$. The commutators $[A, H_{T_i}]$ for each case are analyzed.\\

\subsubsection*{(a) Linear Coupling $H_{T_1}$}

\[
H_{T_1} = \hbar \int d\omega \left( \kappa_1 A^\dagger a_1(\omega) + \kappa_1^* a_1^\dagger(\omega) A \right)
\]
\[
[A, H_{T_1}] = \hbar \int d\omega \left( \kappa_1 [A, A^\dagger a_1(\omega)] + \kappa_1^* [A, a_1^\dagger(\omega) A] \right)
\]
\noindent
$A$ and $a_1(\omega)$ act on different systems (system vs bath), so they commute:

\[
[A, a_1(\omega)] = [A, a_1^\dagger(\omega)] = 0
\]
\noindent
Therefore:

\[
[A, A^\dagger a_1(\omega)] = [A, A^\dagger] a_1(\omega) = a_1(\omega)
\]

\[
[A, a_1^\dagger(\omega) A] = a_1^\dagger(\omega) [A, A] = 0
\]
\noindent
Hence:

\begin{align}
[A, H_{T_1}] = \hbar \int d\omega \, \kappa_1 a_1(\omega)
\end{align}
\noindent
The system operator $A$ experiences the input from bath 1 linearly through the operator $a_1(\omega)$ with strength $\kappa_1$.\\
\subsubsection*{(b) Linear Coupling $H_{T_2}$}

\[
H_{T_2} = \hbar \int d\omega \left( \kappa_2 A^\dagger a_2(\omega) + \kappa_2^* a_2^\dagger(\omega) A \right)
\]

\[
[A, H_{T_2}] = \hbar \int d\omega \left( \kappa_2 [A, A^\dagger a_2(\omega)] + \kappa_2^* [A, a_2^\dagger(\omega) A] \right)
\]

\[
[A, A^\dagger a_2(\omega)] = [A, A^\dagger] a_2(\omega) =  a_2(\omega)
\]

\[
[A, a_2^\dagger(\omega) A] = 0
\]
\noindent
Therefore:

\begin{align}
[A, H_{T_2}] = \hbar \int d\omega \, \kappa_2 a_2(\omega)
\end{align}
\noindent
Again, this is a linear coupling from bath 2 into the system.\\

\subsubsection*{(c) Nonlinear Coupling $H_{T_3}$}

\[
H_{T_3} = \hbar \int d\omega \left( \kappa_3 A^\dagger A^\dagger a_3(\omega) + \kappa_3^* a_3^\dagger(\omega) A A \right)
\]

\[
[A, H_{T_3}] = \hbar \int d\omega \left( \kappa_3 [A, A^\dagger A^\dagger a_3(\omega)] + \kappa_3^* [A, a_3^\dagger(\omega) A A] \right)
\]
\noindent
Since $[A, a_3^\dagger(\omega)] = 0$, the second commutator is zero. So the expression is:

\[
[A, A^\dagger A^\dagger a_3(\omega)] = [A, A^\dagger A^\dagger] a_3(\omega)
\]
\noindent
Using the identity $[A, BC] = [A, B]C + B[A, C]$, one has:

\[
[A, A^\dagger A^\dagger] = [A, A^\dagger] A^\dagger + A^\dagger [A, A^\dagger] = A^\dagger + A^\dagger = 2A^\dagger
\]
\noindent
Thus:
\[
[A, A^\dagger A^\dagger a_3(\omega)] = 2A^\dagger a_3(\omega)
\]
\noindent
Finally:
\begin{align}
[A, H_{T_3}] = \hbar \int d\omega \, 2\kappa_3 A^\dagger a_3(\omega)
\end{align}
\noindent
 This is a nonlinear input from bath 3. It depends on the system excitation via $A^\dagger$, and the bath input $a_3(\omega)$.
\subsection*{Combining all terms}
\begin{align}
[A, H] &= \hbar\omega_0 A 
        + \hbar K A^\dagger A A 
        + \hbar \int d\omega \, \left( \kappa_1 a_1(\omega) + \kappa_2 a_2(\omega) \right) \notag \\
       &\quad + 2\hbar \int d\omega \, \kappa_3 A^\dagger a_3(\omega)
\end{align}

\noindent
And dividing both sides by $i\hbar$:
\begin{align}
\frac{dA}{dt} &= -i\omega_0 A 
                - i K A^\dagger A A 
                - i \kappa_1 \int d\omega \, a_1(\omega) 
                - i \kappa_2 \int d\omega \, a_2(\omega) \notag \\
              &\quad - 2i \kappa_3 \int d\omega \, A^\dagger a_3(\omega)
              \label{23}
\end{align}
The time evolution of each bath operator in the Heisenberg picture is to be computed:
\begin{align}
\frac{d}{dt}a_i(\omega) = \frac{1}{i\hbar}[a_i(\omega), H]
\end{align}
This is done mode by mode, using the Hamiltonians already given for each interaction.
\subsection*{For \( a_1(\omega) \)}
\[
H_{T1} = \hbar \int d\omega' \left( \kappa_1 A^\dagger a_1(\omega') + \kappa_1^* a_1^\dagger(\omega') A \right)
\]
and
\[
H_{A1} = \hbar \int d\omega' \, \omega' a_1^\dagger(\omega') a_1(\omega')
\]
So the total Hamiltonian involving \( a_1(\omega) \) is:
\begin{align}
H = H_{A1} + H_{T1}
\end{align}
\noindent
The commutator:\\
\\
(i) Free evolution:
\[
[a_1(\omega), H_{A1}] = \hbar \int d\omega' \, \omega' [a_1(\omega), a_1^\dagger(\omega') a_1(\omega')]
\]
\noindent
Using:
\[
[a, a^\dagger b] = [a, a^\dagger] b + a^\dagger [a, b] \Rightarrow [a, a^\dagger b] = \delta a
\]
\noindent
So:
\[
[a_1(\omega), a_1^\dagger(\omega') a_1(\omega')] = \delta(\omega - \omega') a_1(\omega') \Rightarrow [a_1(\omega), H_{A1}] = \hbar \omega a_1(\omega)
\]
\noindent
(ii) Interaction term:
\[
[a_1(\omega), H_{T1}] = [a_1(\omega), \hbar \int d\omega' \left( \kappa_1 A^\dagger a_1(\omega') + \kappa_1^* a_1^\dagger(\omega') A \right)]
\]
\noindent
Only the second term has a non-zero commutator:
\begin{align}
[a_1(\omega), a_1^\dagger(\omega')] &= \delta(\omega - \omega') \notag \\
\Rightarrow [a_1(\omega), H_{T1}] &= \hbar \int d\omega' \, \kappa_1^* \delta(\omega - \omega') A = \hbar \kappa_1^* A
\end{align}
\noindent
Therefore:
\begin{align}
\frac{d}{dt} a_1(\omega) = \frac{1}{i\hbar} [a_1(\omega), H] = -i\omega a_1(\omega) - i \kappa_1^* A 
\end{align}
\noindent
\subsection*{For \( a_2(\omega) \)}
\noindent
Similarly:
\[
H_{T2} = \hbar \int d\omega \left( \kappa_2 A^\dagger a_2(\omega) + \kappa_2^* a_2^\dagger(\omega) A \right)
\]
\begin{align}
\frac{d}{dt} a_2(\omega) = -i\omega a_2(\omega) - i \kappa_2^* A \
\end{align}
\noindent
\subsection*{For \( a_3(\omega) \)}
\noindent
This bath is nonlinearly coupled:
\[
H_{T3} = \hbar \int d\omega \left( \kappa_3 A^\dagger A^\dagger a_3(\omega) + \kappa_3^* a_3^\dagger(\omega) A A \right)
\]

\[
[a_3(\omega), H_{T3}] = [a_3(\omega), \hbar \int d\omega' \left( \kappa_3 A^\dagger A^\dagger a_3(\omega') + \kappa_3^* a_3^\dagger(\omega') A A \right)]
\]
\noindent
Only the second term contributes:
\[
[a_3(\omega), a_3^\dagger(\omega')] = \delta(\omega - \omega') \Rightarrow [a_3(\omega), H_{T3}] = \hbar \kappa_3^* A A
\]
\noindent
The free part again gives:
\[
[a_3(\omega), H_{A3}] = \hbar \omega a_3(\omega)
\]
\noindent
So the final term is:
\begin{align}
\frac{d}{dt} a_3(\omega) = -i\omega a_3(\omega) - i \kappa_3^* A A 
\end{align}

\section*{Quantum Langevin Equation Derivation}
\noindent
The aim is to derive the Quantum Langevin equation for a system operator \( A(t) \) linearly coupled to a bosonic bath of harmonic oscillators. This equation will include both a damping term and a noise (input) term.

\subsection*{Heisenberg Equation for Bath Operator \( a(\omega, t) \)}
\noindent
From the total Hamiltonian, the equation of motion for the bath operator is:
\begin{align}
\frac{d}{dt} a(\omega, t) = -i \omega a(\omega, t) - i \kappa^* A(t)
\end{align}

\subsection*{Solution via Integrating Factor}
\noindent
Multiplying by \( e^{i\omega t} \):
\[
e^{i\omega t} \frac{d}{dt} a(\omega, t) = -i\omega a(\omega, t) e^{i\omega t} - i \kappa^* A(t) e^{i\omega t}
\]
\[
\Rightarrow \frac{d}{dt} \left[ a(\omega, t) e^{i\omega t} \right] = -i \kappa^* A(t) e^{i\omega t}
\]
\noindent
Integrating from \( t_0 \) to \( t \):
\[
a(\omega, t) e^{i\omega t} - a(\omega, t_0) e^{i\omega t_0} = -i \kappa^* \int_{t_0}^t d\tau\, A(\tau) e^{i\omega \tau}
\]
\noindent
Multiplying by \( e^{-i\omega t} \):
\begin{align}
a(\omega, t) = a(\omega, t_0) e^{-i\omega(t - t_0)} - i \kappa^* \int_{t_0}^t d\tau\, A(\tau) e^{-i\omega(t - \tau)} 
\label{32}
\end{align}

\subsection*{Equation of Motion for the System Operator \( A(t) \)}

\begin{align}
\frac{dA}{dt} &= -i\omega_0 A 
               - i K A^\dagger A A 
               - i \kappa_1 \int d\omega\, a_1(\omega) 
               - i \kappa_2 \int d\omega\, a_2(\omega) \notag \\
             &\quad - 2i \kappa_3 \int d\omega\, A^\dagger a_3(\omega)
             \label{33}
\end{align}
\noindent
Substituting Eq \eqref{32} in Eq \eqref{33}:
\begin{align}
\frac{dA(t)}{dt} &= \dots 
               - i \int d\omega\, \kappa\, a(\omega, t_0) e^{-i\omega(t - t_0)} \notag \\
             &\quad + \int d\omega\, |\kappa|^2 \int_{t_0}^t d\tau\, A(\tau) e^{-i\omega(t - \tau)}
\end{align}
\noindent
The input field is defined \( a_{\text{in}}(t) \) as:
\[
a_{\text{in}}(t) \equiv \frac{1}{2\pi} \int d\omega\, a(\omega, t_0) e^{-i\omega(t - t_0)} 
\]
\noindent
Then Eq.~(33) becomes:
\begin{align}
\frac{dA(t)}{dt} = \dots  + \int d\omega\, |\kappa|^2 \int_{t_0}^t d\tau\, A(\tau) e^{-i\omega(t - \tau)} - i \kappa \cdot 2\pi a_{\text{in}}(t)
\end{align}

\subsection*{Evaluating the Damping Integral (Markov Approximation)}
\noindent
Assuming a broadband bath:
\begin{align}
\int d\omega\, e^{-i\omega(t - \tau)} = 2\pi \delta(t - \tau)
\end{align}
\noindent
Thus:
\[
\int d\omega\, |\kappa|^2 \int_{t_0}^t d\tau\, A(\tau) e^{-i\omega(t - \tau)} \approx |\kappa|^2 \cdot 2\pi A(t)
\]
\noindent
where :
\[ \gamma \equiv 2\pi |\kappa|^2 \Rightarrow |\kappa| = \sqrt{\frac{\gamma}{2\pi}}
\]

\noindent
then:
\begin{align}
\frac{dA(t)}{dt} = \dots - \gamma A(t) - i \cdot 2\pi \kappa a_{\text{in}}(t)   
\end{align}

\noindent
Using \( \kappa = \sqrt{\frac{\gamma}{2\pi}} e^{i\phi} \):
\[
2\pi \kappa = \sqrt{2\pi\gamma} e^{i\phi} 
\]

\subsection*{Final Langevin Equation}

\begin{equation}
\boxed{
\frac{dA(t)}{dt} = \dots - \gamma A(t) - i \cdot \sqrt{2\gamma} e^{i\phi} a_{\text{in}}(t)}
\label{eq:langevin}
\end{equation}

\noindent
This is the Langevin equation with:
\begin{itemize}
    \item \textbf{Damping term:} \( -\gamma A(t) \)
    \item \textbf{Noise term:} \( -i \cdot \sqrt{2\gamma} e^{i\phi} a_{\text{in}}(t) \)
\end{itemize}
\noindent
This gives us terms for the first two bath modes :
\begin{table}[h!]
\centering
\begin{tabular}{@{}ll@{}}
\toprule
\textbf{Interaction term} & \textbf{Langevin contribution} \\
\midrule
$-i \kappa_1 \int d\omega \, a_1(\omega)$ & $-\gamma_1 A(t) - i \sqrt{2\gamma_1} e^{i\phi_1} a_{\text{in},1}(t)$ \\
$-i \kappa_2 \int d\omega \, a_2(\omega)$ & $-\gamma_2 A(t) - i \sqrt{2\gamma_2} e^{i\phi_2} a_{\text{in},2}(t)$ \\
\bottomrule
\end{tabular}
\caption{Langevin contributions from coupling to two bosonic baths.}
\end{table}

\noindent
Similarly, one can derive the third term and consequently put all terms together to get the final Langevin equation coupled to 3 different modes. 
\begin{align}
- \gamma_3 A^\dagger A A - i 2 \sqrt{\gamma_3} e^{i\phi_3} A^\dagger a^{\text{in}}_3(t)
\end{align}

\begin{align}
\frac{dA}{dt} = 
& -i\omega_0 A 
- iK A^\dagger A A 
- \gamma A 
- \gamma_3 A^\dagger A A \notag \\
& - i \sqrt{2\gamma_1} e^{i\phi_1} a_1^{\text{in}}(t) 
- i \sqrt{2\gamma_2} e^{i\phi_2} a_2^{\text{in}}(t) \notag \\
& - i 2 \sqrt{\gamma_3} e^{i\phi_3} A^\dagger a_3^{\text{in}}(t)
\end{align} where \begin{equation}
\gamma_1 + \gamma_2 = \gamma
\label{eq:gamma_sum}
\end{equation}
\noindent
Next, the relations between the outgoing bath modes, incoming bath modes and the cavity mode A are defined. \\
\noindent
Free bath evolution:
\begin{equation}
    a_i(\omega, t) = a_i(\omega, t_0) e^{-i\omega(t - t_0)}
\end{equation}
\noindent
Input field:
\begin{equation}
    a_{\text{in},i}(t) = \frac{1}{2\pi} \int d\omega\, a_i(\omega, t_0) e^{-i\omega(t - t_0)}
\end{equation}
\noindent
Output field:
\begin{equation}
    a_{\text{out},i}(t) = \frac{1}{2\pi} \int d\omega\, a_i(\omega, t_1) e^{-i\omega(t - t_1)}
\end{equation}
\noindent
Solving the Bath Operator EOM : \\
\noindent
From the total Hamiltonian, the Heisenberg equation for bath mode \( i \) is:
\begin{equation}
    \frac{d}{dt} a_i(\omega, t) = -i\omega a_i(\omega, t) -i \kappa_i A(t)
\end{equation}
This has the solution:
\begin{equation}
    a_i(\omega, t) = a_i(\omega, t_0) e^{-i\omega(t - t_0)} -i \kappa_i \int_{t_0}^{t} d\tau\, A(\tau) e^{-i\omega(t - \tau)}
\end{equation}
\noindent
Boundary Condition for Mode 1\\
\noindent
Using:
\[
H_{\text{int},1} = -i\kappa_1 \int d\omega \left( A^\dagger a_1(\omega) - a_1^\dagger(\omega) A \right)
\]
\begin{equation}
    \frac{d}{dt} a_1(\omega, t) = -i\omega a_1(\omega, t) + \kappa_1 A(t)
\end{equation}
\begin{equation}
    a_1(\omega, t_1) = a_1(\omega, t_0) e^{-i\omega(t_1 - t_0)} +-i \kappa_1 \int_{t_0}^{t_1} d\tau\, A(\tau) e^{-i\omega(t_1 - \tau)}
\end{equation}
\noindent
Then plug into the output field:
\begin{align}
a_{\text{out},1}(t) 
&= \frac{1}{2\pi} \int d\omega\, 
\bigg[ a_1(\omega, t_0) e^{-i\omega(t_1 - t_0)} e^{-i\omega(t - t_1)} \notag \\
&\quad -i \kappa_1 \int_{t_0}^{t_1} d\tau\, A(\tau) 
e^{-i\omega(t_1 - \tau)} e^{-i\omega(t - t_1)} \bigg] \notag \\
&= \frac{1}{2\pi} \int d\omega\, a_1(\omega, t_0) e^{-i\omega(t - t_0)} \notag \\
&\quad -i \kappa_1 \int_{t_0}^{t_1} d\tau\, A(\tau) 
\left( \frac{1}{2\pi} \int d\omega\, e^{-i\omega(t - \tau)} \right)
\end{align}

\begin{equation}
a_{\text{out},1}(t) = a_{\text{in},1}(t) - i\, \sqrt{2\gamma_1} e^{-i\phi_1} A(t)
\label{eq:io1}
\end{equation}
\noindent
where:
\[
\sqrt\frac{\gamma_1}{\pi} e^{i\phi_1} = \kappa_1
\]
\begin{equation}
\boxed{
a_{\text{out},1}(t) - a_{\text{in},1}(t) = -i\, \sqrt{2\gamma_1} e^{-i\phi_1} A(t) \quad 
}
\end{equation}
\noindent
Mode 2 has an identical derivation following a similar procedure:
\[
a_{\text{out},2}(t) = a_{\text{in},2}(t) - i\, \sqrt{2\gamma_2} e^{-i\phi_2} A(t)
\]
\begin{equation}
\boxed{
a_{\text{out},2}(t) - a_{\text{in},2}(t) = -i\, \sqrt{2\gamma_2} e^{-i\phi_2} A(t) \quad 
}
\label{eq:io2}
\end{equation}
\noindent
Mode 3 — Nonlinear Coupling)
\\
\noindent
Interaction Hamiltonian:
\[
H_{\text{int},3} = -i 2\kappa_3 \int d\omega\, A^\dagger a_3(\omega)
\]
Heisenberg EOM:
\begin{equation}
    \frac{d}{dt} a_3(\omega, t) = -i\omega a_3(\omega, t) + 2\kappa_3 A^\dagger(t)
\end{equation}
\begin{equation}
    a_3(\omega, t) = a_3(\omega, t_0) e^{-i\omega(t - t_0)} + 2\kappa_3 \int_{t_0}^{t} d\tau\, A^\dagger(\tau) e^{-i\omega(t - \tau)}
\end{equation}
\noindent
Then the output field:
\[
a_{\text{out},3}(t) = a_{\text{in},3}(t) - i\, \sqrt\gamma_3 e^{-i\phi_3} A(t) A(t)
\]
\begin{equation}
\boxed{
a_{\text{out},3}(t) - a_{\text{in},3}(t) = -i\, \sqrt\gamma_3 e^{-i\phi_3} A(t)^2 \quad 
}
\label{eq:io3}
\end{equation}
\noindent
Finally,
\begin{align}
    a_{\text{out},1}(t) - a_{\text{in},1}(t) &= -i\, \sqrt{2\gamma_1} e^{-i\phi_1} A(t) \notag \\
    a_{\text{out},2}(t) - a_{\text{in},2}(t) &= -i\, \sqrt{2\gamma_2} e^{-i\phi_2} A(t) \notag \\
    a_{\text{out},3}(t) - a_{\text{in},3}(t) &= -i\, \sqrt\gamma_3 e^{-i\phi_3} A(t)^2 \notag
\end{align}
To determine the device's response to a classical pump without any signal or noise, the incoming noise terms are set to zero. Assumptions:
\begin{align}
a_{\text{in},2} &= 0, \quad a_{\text{in},3} = 0  \\
a_{\text{in},1}(t) &= b_{\text{in},1} e^{-i(\omega_p t + \psi_1)} \quad \text{where } b_{\text{in},1} \in \mathbb{R} \quad 
\end{align}
\noindent
Here, $\psi_1$ is the phase  of the pump. Ansatz for cavity field:
\begin{equation}
A(t) = B e^{-i(\omega_p t + \phi_B)} \quad \text{with } B \in \mathbb{R}_+, \ \phi_B \in \mathbb{R} 
\end{equation}

\noindent
Input-Output Relation at Port 1 (Eq.~\eqref{eq:io1})
\begin{equation}
a_{\text{out},1}(t) - a_{\text{in},1}(t) = -i \sqrt{2\gamma_1} e^{-i\phi_1} A(t)
\end{equation}
\noindent
Substituting ansatz into the EOM  (Eq.~\eqref{eq:EOM_A})

\noindent
From the ansatz:
\[
A(t) = B e^{-i(\omega_p t + \phi_B)}, \quad A^\dagger(t) = B e^{i(\omega_p t + \phi_B)}
\]
\[
A^\dagger A A = B^3 e^{-i(\omega_p t + \phi_B)}
\]
\[
\frac{dA}{dt} = -i \omega_p B e^{-i(\omega_p t + \phi_B)} = -i \omega_p A(t)
\]
\noindent
Plugging into the EOM (from Eq.~\eqref{23} under assumptions \( a_{\text{in},2} = a_{\text{in},3} = 0 \)):
\[
-i \omega_p A = -i \omega_0 A - i K A^\dagger A A - \gamma A - \gamma_3 A^\dagger A A - i \sqrt{2\gamma_1} e^{i\phi_1} a_{\text{in},1}(t)
\]
\noindent
Bring all terms to one side:
\[
[i(\omega_0 - \omega_p) + \gamma] A + (iK + \gamma_3) |A|^2 A = -i \sqrt{2\gamma_1} e^{i\phi_1} a_{\text{in},1}(t)
\]
\noindent
Substituting:
\[
A = B e^{-i(\omega_p t + \phi_B)}, \quad |A|^2 = B^2, \quad a_{\text{in},1}(t) = b_{\text{in},1} e^{-i(\omega_p t + \psi_1)}
\]
\noindent
LHS:
\[
[i(\omega_0 - \omega_p) + \gamma + (iK + \gamma_3) B^2] B e^{-i(\omega_p t + \phi_B)}
\]
\noindent
RHS:
\[
-i \sqrt{2\gamma_1} b_{\text{in},1} e^{i\phi_1} e^{-i(\omega_p t + \psi_1)}
\]
\noindent
Equating both sides:
\begin{equation}
[i(\omega_0 - \omega_p) + \gamma + (iK + \gamma_3) B^2] B = -i \sqrt{2\gamma_1} b_{\text{in},1} e^{i(\phi_1 + \phi_B - \psi_1)}
\end{equation}
\noindent
Deriving output field
\noindent
from the input-output relation:
\[
a_{\text{out},1}(t) = a_{\text{in},1}(t) - i \sqrt{2\gamma_1} e^{-i\phi_1} A(t)
\]
\noindent
Substituting $ a_{\text{out},1}(t)= b_{\text{out},1} e^{-i(\omega_p t + \psi_1)}$ :
\[
b_{\text{out},1} e^{-i(\omega_p t + \psi_1)} = b_{\text{in},1} e^{-i(\omega_p t + \psi_1)} - i \sqrt{2\gamma_1} e^{-i\phi_1} B e^{-i(\omega_p t + \phi_B)}
\]
\noindent
Finally,
\begin{equation}
b_{\text{out},1} = b_{\text{in},1} - i \sqrt{2\gamma_1} B e^{-i(\phi_1 + \phi_B - \psi_1)} 
\end{equation}
\noindent
\subsection*{Derivation of the cubic equation}
\noindent
Given:
\begin{equation}
\left[i(\omega_0 - \omega_p) + \gamma\right] B + (iK + \gamma_3) B^3 = -i 2\gamma_1 b_{\text{in},1} e^{i(\phi_1 + \phi_B - \psi_1)}
\end{equation}
\noindent
Defining:
\[
E = |B|^2 = B B^*, \quad \text{so that} \quad |B^3|^2 = E^3
\]

\[
A = i(\omega_0 - \omega_p) + \gamma, \quad C = iK + \gamma_3, \quad F = - i 2\gamma_1 b_{\text{in},1} e^{i(\phi_1 + \phi_B - \psi_1)}
\]
\noindent
Then it becomes:
\[
AB + C B^3 = F
\]
\noindent
Taking the complex conjugate:
\[
A^* B^* + C^* (B^*)^3 = F^*
\]
\noindent
Multiplying the equation by its complex conjugate:
\[
(AB + C B^3)(A^* B^* + C^* (B^*)^3) = F F^*
\]
\noindent
Expanding the left-hand side:
\[
|A|^2 |B|^2 + A C^* |B|^2 (B^*)^2 + A^* C |B|^2 B^2 + |C|^2 |B|^6
\]
\noindent
Factoring using \( E = |B|^2 \) and \( B^2 = E e^{2i\theta}, \, (B^*)^2 = E e^{-2i\theta} \):
\[
|A|^2 E + A C^* E^2 e^{-2i\theta} + A^* C E^2 e^{2i\theta} + |C|^2 E^3
\]
\noindent
And grouping the complex terms:
\[
|A|^2 E + 2E^2 \Re(A C^*) + |C|^2 E^3 = |F|^2
\]
\noindent
The coefficients can be computed explicitly:
\begin{align*}
|A|^2 &= \gamma^2 + (\omega_0 - \omega_p)^2 \\
|C|^2 &= \gamma_3^2 + K^2 \\
\Re(A C^*) &= \gamma \gamma_3 + (\omega_0 - \omega_p) K \\
|F|^2 &= (2\gamma_1 b_{\text{in},1})^2 = 4\gamma_1^2 (b_{\text{in},1})^2
\end{align*}
\noindent
Substituting into the equation:
\begin{align}
E (\gamma^2 + (\omega_0 - \omega_p)^2) 
&+ 2E^2 \left[\gamma \gamma_3 + (\omega_0 - \omega_p) K \right] \notag \\
&+ E^3 (\gamma_3^2 + K^2) = 4\gamma_1^2 (b_{\text{in},1})^2
\end{align}
\noindent
Dividing through by \( \gamma_3^2 + K^2 \) to get a normalized cubic equation:
\[
E^3 + 2 \frac{(\omega_0 - \omega_p) K + \gamma \gamma_3}{K^2 + \gamma_3^2} E^2 + \frac{(\omega_0 - \omega_p)^2 + \gamma^2}{K^2 + \gamma_3^2} E = \frac{4\gamma_1^2 (b_{\text{in},1})^2}{K^2 + \gamma_3^2}
\]
\noindent
Rewriting in standard form:
\begin{align}
E^3 + 2 \frac{(\omega_0 - \omega_p) K + \gamma \gamma_3}{K^2 + \gamma_3^2} E^2 + \frac{(\omega_0 - \omega_p)^2 + \gamma^2}{K^2 + \gamma_3^2} E - \frac{4\gamma_1^2 (b_{\text{in},1})^2}{K^2 + \gamma_3^2} = 0
\end{align}
\subsection*{Derivation of \(\frac{dE}{d\omega_p}\) and Resonance Peak Condition}
\noindent
One can start from the cubic equation for the energy \( E = |B|^2 \) of the form:

\[
E^3 + \frac{2[(\omega_0 - \omega_p)K + \gamma \gamma_3]}{K^2 + \gamma_3^2} E^2 + \frac{(\omega_0 - \omega_p)^2 + \gamma^2}{K^2 + \gamma_3^2} E - \frac{2 \gamma_1 |b_{\text{in},1}|^2}{K^2 + \gamma_3^2} = 0 
\]
\noindent
For the sake of simplicity, defining:
\[
\Delta = \omega_0 - \omega_p, \quad \beta = K^2 + \gamma_3^2
\]
\noindent
Then the equation becomes:
\[
f(E, \omega_p) = E^3 + \frac{2(K \Delta  + \gamma \gamma_3)}{\beta} E^2 + \frac{\Delta^2 + \gamma^2}{\beta} E - \frac{2 \gamma_1 |b_{\text{in},1}|^2}{\beta} = 0
\]
\noindent
To find the condition for the resonance peak,  differentiating implicitly with respect to \(\omega_p\):

\[
\frac{d f}{d \omega_p} = \frac{\partial f}{\partial E} \cdot \frac{dE}{d\omega_p} + \frac{\partial f}{\partial \omega_p} = 0
\Rightarrow
\frac{dE}{d\omega_p} = -\frac{\frac{\partial f}{\partial \omega_p}}{\frac{\partial f}{\partial E}} 
\]
\noindent
Computing the partial derivatives:

\paragraph{Derivative with respect to \( E \)}

\[
\frac{\partial f}{\partial E} = 3E^2 + \frac{4(\Delta K + \gamma \gamma_3)}{\beta} E + \frac{\Delta^2 + \gamma^2}{\beta}
\]

\paragraph{Derivative with respect to \( \omega_p \)}
Since \( \frac{d\Delta}{d\omega_p} = -1 \), one has
\[
\frac{\partial f}{\partial \omega_p} = 
\frac{2K(-1)}{\beta} E^2 + \frac{2\Delta(-1)}{\beta} E = 
-\frac{2K}{\beta} E^2 - \frac{2\Delta}{\beta} E
\]

\paragraph{Substituting}

\[
\frac{dE}{d\omega_p} = \frac{2K E^2 + 2\Delta E}{\beta \left[3E^2 + \dfrac{4(\Delta K + \gamma \gamma_3)}{\beta} E + \dfrac{\Delta^2 + \gamma^2}{\beta} \right]}
\]
Factoring the numerator:

\begin{align}
\frac{dE}{d\omega_p} = \frac{2E (K E + \Delta)}{\beta \left[3E^2 + \dfrac{4(\Delta K + \gamma \gamma_3)}{\beta} E + \dfrac{\Delta^2 + \gamma^2}{\beta} \right]} 
\label{65}
\end{align}
\paragraph{Condition for Maximum Response}
Setting \( \frac{dE}{d\omega_p} = 0 \). This occurs when the numerator of Eq \eqref{65} is zero:

\[
2E (K E + \Delta) = 0
\Rightarrow K E + \Delta = 0
\Rightarrow \boxed{\omega_0 - \omega_p + K E = 0} 
\]
\noindent
This condition shows that the resonance peak is shifted by the Kerr nonlinearity term \( K E \), consistent with the behavior of Duffing-type nonlinear oscillators.

\subsection*{Calculating \(\frac{d\omega_p}{dE}\) and \(\frac{d^2\omega_p}{dE^2}\) } 
\noindent
To find the condition for the resonance peak, differentiating implicitly with respect to $dE$:

\[
\frac{d f}{dE} = \frac{\partial f}{\partial \omega_p} \cdot \frac{d\omega_p}{dE} + \frac{\partial f}{\partial E} = 0
\Rightarrow
\frac{d \omega_p}{d E} = -\frac{\frac{\partial f}{\partial E}}{\frac{\partial f}{\partial \omega_p}} 
\]
\noindent
Computing the partial derivatives:

\paragraph{Derivative with respect to \( E \)}

\[
\frac{\partial f}{\partial E} = 3E^2 + \frac{4(\Delta K + \gamma \gamma_3)}{\beta} E + \frac{\Delta^2 + \gamma^2}{\beta}
\]

\paragraph{Derivative with respect to \( \omega_p \)}

Since \( \frac{d\Delta}{d\omega_p} = -1 \), one obtains:

\[
\frac{\partial f}{\partial \omega_p} = 
\frac{2K(-1)}{\beta} E^2 + \frac{2\Delta(-1)}{\beta} E = 
-\frac{2K}{\beta} E^2 - \frac{2\Delta}{\beta} E
\] 

\begin{align}
\frac{d \omega_p}{dE} = \frac{3E^2\beta + 4(\Delta K + \gamma \gamma_3)E + \Delta^2+ \gamma^2}{E^2 [K+2(\omega_0-\omega_p)]}
\end{align}
\noindent
In a nonlinear driven resonator, the intracavity field amplitude \( E \) depends on the driving (pump) frequency \( \omega_p \). The system exhibits bistability, meaning that for a given drive frequency, there may be two stable values of \( E \). The transitions between these bistable states (i.e., switching points) occur at the turning points of the response curve \( E(\omega_p) \). These are precisely the points where the slope of the response diverges, i.e., the derivative of the drive frequency with respect to amplitude vanishes.

\begin{equation}
\frac{\partial \omega_p}{\partial E} = 0
\label{eq:crit_slope}
\end{equation}

\noindent
Setting the numerator of $\frac{\partial \omega_p}{\partial E}$ to 0, one obtains :

\begin{equation}
3E^2K^2 + 3E^2\gamma_3^2 + 4EK(\omega_0 - \omega_p) +4\gamma\gamma_3E+(\omega_0-\omega_p)^2+\gamma^2 = 0
\end{equation}

\begin{equation}
\left( \gamma + 2\gamma_3 E \right)^2 = (K^2 + \gamma_3^2) E^2 - (\omega_0 - \omega_p + 2K E)^2
\label{70}
\end{equation}
\noindent
Similarly for $\frac{\partial^2 \omega_p}{\partial E^2}$ 

\begin{equation}
6(K^2 + \gamma_3^2)E + 4\left[ (\omega_0 - \omega_p)K + \gamma \gamma_3 \right] = 0
\label{71}
\end{equation}
\noindent
When both equations Eq.~\eqref{70} and ~\eqref{71} are satisfied, the two instable points merge into one — the critical point of the Duffing-type nonlinear response. At or near this critical point, the slope \( \partial E / \partial \omega_p \) becomes very large or diverges, which corresponds to a situation of large parametric gain. While large gain is desirable for amplification, operating exactly at the critical point can make the system unstable due to strong sensitivity to perturbation. Therefore, in practice, one aims to design the parametric amplifier such that it operates close to but not inside the bistable region. This ensures high gain while maintaining dynamical stability.
\noindent
Condition for Critical Point: To ensure that a critical point exists (i.e., a point where both derivatives vanish), the following condition must be satisfied:

\begin{equation}
|K| > \sqrt{3} \gamma_3
\label{eq:critical_K}
\end{equation}

\noindent
This implies that the Kerr nonlinearity \( K \) must be strong enough compared to the two-photon loss rate \( \gamma_3 \). If this condition is not satisfied, the system cannot exhibit bistability. At the critical point, the amplitude of the resonator field reaches the value:
\begin{equation}
    E_c = \frac{2\gamma}{\sqrt{3} (|K| - \sqrt{3} \gamma_3)}
\end{equation}
\noindent
This value diverges as \( \gamma_3 \to |K|/\sqrt{3} \), consistent with the disappearance of bistability at that limit.
Critical detuning is defined as the detuning between the cavity resonance \( \omega_0 \) and the pump frequency \( \omega_p \) at the critical point given by:

\begin{equation}
\omega_0 - \omega_p = -\frac{\gamma}{K} \cdot \frac{|K|}{K}
\left[
4 \gamma_3 |K| + \frac{\sqrt{3}(K^2 + \gamma_3^2)}{K^2 - 3 \gamma_3^2}
\right]
\label{eq:critical_detuning}
\end{equation}

\noindent
This expression ensures that both the first and second derivatives of \( \omega_p(E) \) vanish at the critical point. The input pump amplitude squared to reach the critical point is the following.
\begin{equation}
|b_{\text{in},1c}|^2 = \frac{4}{3\sqrt{3}} \cdot \frac{\gamma^3 (K^2 + \gamma_3^2)}{\gamma_1 (|K| - \sqrt{3} \gamma_3)^3}
\label{eq:critical_input}
\end{equation}

\noindent
This is the minimum input power required to drive the system into the bistable regime. As \( \gamma_3 \) increases, the power required increases dramatically. Large parametric gain occurs near the critical point, where the system's response is highly sensitive to small changes in drive frequency or amplitude. To avoid instabilities due to bistability, it is desirable to operate close to—but not within—the bistable regime. When \( \gamma_3 > |K|/\sqrt{3} \), the bistable region becomes inaccessible, and the system behaves in a monostable manner. The presence of two-photon loss \( \gamma_3 \) modifies the shape and position of the Duffing response curve. 

\section{Linearization}
The signals entering the input port and the noise entering the loss ports are considered small as compared to the pump.  
\begin{equation}
E^3 + 2\left[\frac{(\omega_0 - \omega_p)}{K}\right] E^2 + \left[\frac{(\omega_0 - \omega_p)^2 + \gamma^2}{K^2}\right] E - \frac{2\gamma}{K^2} \left(b_{\text{in},1}\right)^2 = 0
\label{76}
\end{equation}
\\
where the incoming pump amplitude for operation at the critical point is given by
\begin{equation}
\left(b_{\text{in}, 1c}\right)^2 = \frac{4}{3\sqrt{3}} \frac{\gamma^2}{|K|}
\end{equation}
solving equation Eq.~\eqref{76} for $b_{in,1}$= $0.5$$b_{in,1c}$ , $\omega_0$ = $100$$e^9$ , $K$ = -$9.99e^{-5}$$\omega_0$ and $\gamma$ = $0.0032$ $\omega_0$ yields the plots as shown in figure 2  (where the detuning ratio $y =$ $\omega_0$-$\omega_p$/$\omega_0$).\\
\\
\noindent
The Heisenberg-Langevin equation is given by:
\begin{align}
\frac{d a}{dt} + W a + V a^\dagger = F(t)
\end{align}
here 
\begin{itemize}
    \item \(W\) is a complex damping parameter,
    \item \(V\) is a complex gain (or squeezing) parameter,
    \item \(F(t)\) is the driving noise/input operator.
\end{itemize}
\noindent
\\
E= B$^2$ and
\begin{align}
W = i(\omega_0 - \omega_p) + \gamma + 2iKB^2
\end{align}
\begin{align}
V = iKB^2 e^{-2i\phi_B}
\end{align}
\begin{align}
F = -i\sqrt{2\gamma_1} e^{i\phi_1} c_{\text{in},1}
\end{align}
\begin{figure}[H]
    \centering
    \begin{minipage}[b]{0.5\textwidth}
        \centering
        \includegraphics[width=\textwidth]{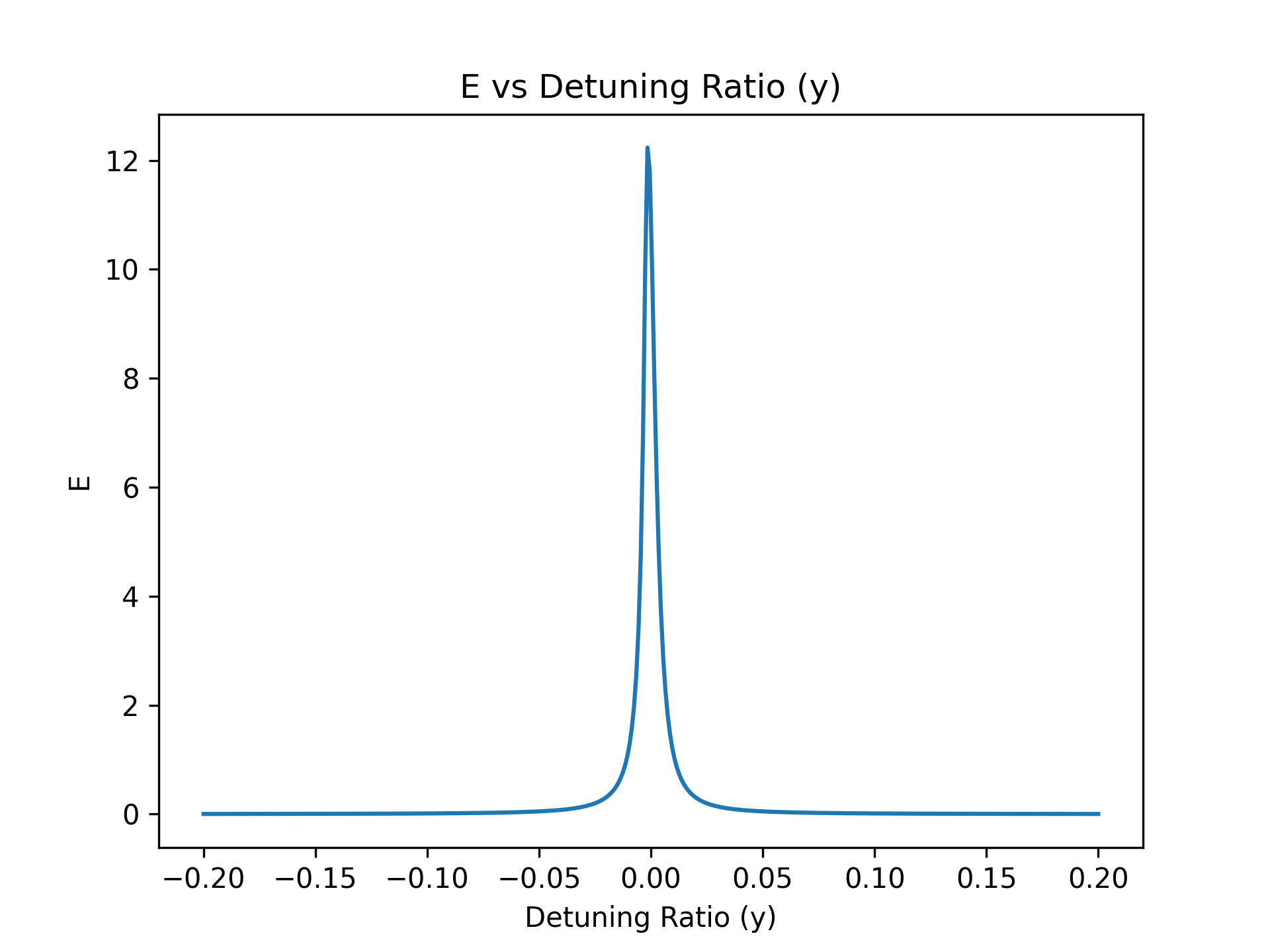}
        \label{fig:figure1}
    \end{minipage}
    \begin{minipage}[b]{0.5\textwidth}
        \centering
        \includegraphics[width=\textwidth]{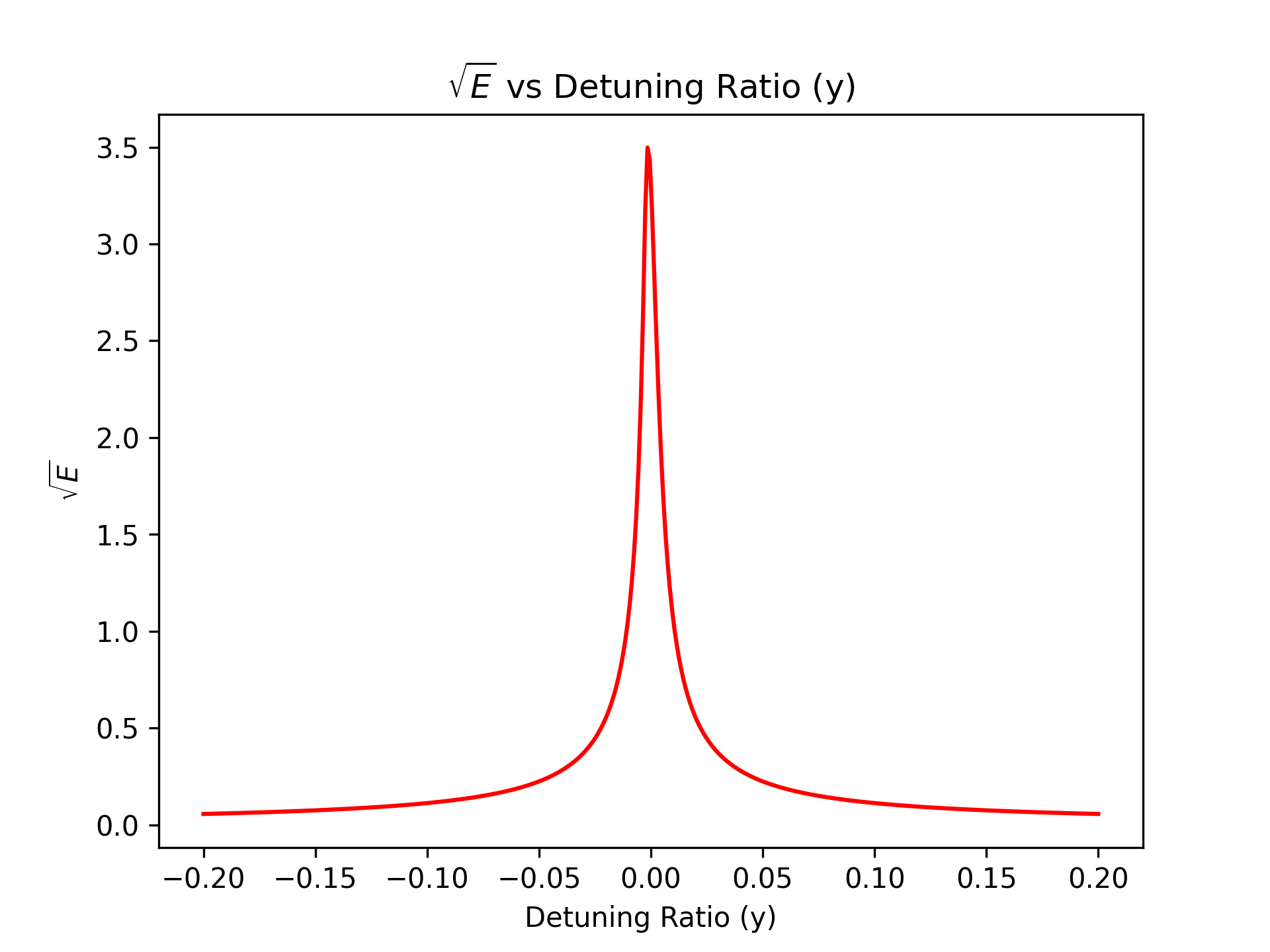}
        \label{fig:figure2}
    \end{minipage}
    \label{e}
    \caption{E vs Detuning Ratio and $\sqrt{E}$ vs Detuning Ratio}
\end{figure}
\noindent
Differentiating the linearized equation with respect to time \( t \):
\[
\frac{d^2 a}{dt^2} + W \frac{d a}{dt} + \frac{d W}{dt} a + W \frac{d a}{dt} + \frac{d V}{dt} a^\dagger + V \frac{d a^\dagger}{dt} = \frac{d F}{dt}
\]
Collecting like terms:

\[
\frac{d^2 a}{dt^2} + 2 W \frac{d a}{dt} + \left(\frac{d W}{dt} + \frac{d V}{dt} + W^2 - V^2\right)a = \frac{d F}{dt} + W^* F - V F^\dagger(t)
\]
This reduces to:
\[
\frac{d^2 a}{dt^2} + 2 \Re(W) \frac{d a}{dt} + \left(|W|^2 - |V|^2\right)a = \Gamma(t)
\]
where:
\begin{align}
\Gamma(t) = \frac{d F}{dt} + W^* F - V F^\dagger(t)
\end{align}
Solving the Homogeneous Equation by assuming a solution of the form \( a = e^{-\lambda t} \) :
\[
\frac{d^2 a}{dt^2} + 2 \Re(W) \frac{d a}{dt} + \left(|W|^2 - |V|^2\right)a = 0
\]
Substituting \( a = e^{-\lambda t} \) gives:
\[
\lambda^2 - 2 \Re(W) \lambda + \left(|W|^2 - |V|^2\right) = 0
\]
This is a quadratic equation in \( \lambda \), and its roots are:
\[
\lambda_{0,1} = \Re(W) \pm \sqrt{\Re^2(W) - (|W|^2 + |V|^2)}
\]
Substituting the expressions for \( W \) and \( V \):
\begin{align}
\lambda_0 = \gamma - \sqrt{K^2 B^4 - (\omega_0 - \omega_p + 2K B^2)^2}
\end{align}

\begin{align}
\lambda_1 = \gamma + \sqrt{K^2 B^4 - (\omega_0 - \omega_p + 2K B^2)^2}
\end{align}
\noindent
Critical Slowing Down: When the root \( \lambda_0 \) becomes zero.  
\begin{align}
\Re(W) = \sqrt{\Re^2(W) - (|W|^2 + |V|^2)}
\end{align}
This condition corresponds to critical slowing down, which occurs when the slope of \( E \) with respect to \( \omega_p \) becomes infinite. 

\section*{Fourier Transform of the Time-Domain Equation}
The second-order differential equation in the time domain:
\begin{equation}
\frac{d^2 a(t)}{dt^2} + 2\Re(W)\frac{da(t)}{dt} + \left( |W|^2 - |V|^2 \right) a(t) = \Gamma(t) 
\end{equation}

\noindent
The goal is to transform this into the frequency domain using the Fourier transform.

\subsection*{Fourier Transform Properties}
Let $f(t) \xrightarrow{\mathcal{F}} f(\omega)$ denote the Fourier transform. The following properties are used:

\begin{align}
\mathcal{F}\left[ \frac{df(t)}{dt} \right] &= i\omega f(\omega) \\
\mathcal{F}\left[ \frac{d^2 f(t)}{dt^2} \right] &= -\omega^2 f(\omega) \\
\mathcal{F}[a(t)] &= a(\omega) \\
\mathcal{F}[\Gamma(t)] &= \Gamma(\omega)
\end{align}

\subsection*{Term-by-Term Fourier Transform of Equation}

Apply the Fourier transform to each term:

\begin{itemize}
    \item First term:
    \[
    \mathcal{F} \left[ \frac{d^2 a(t)}{dt^2} \right] = -\omega^2 a(\omega)
    \]

    \item Second term:
    \[
    \mathcal{F} \left[ 2\Re(W)\frac{da(t)}{dt} \right] = 2i\omega\,\Re(W) a(\omega)
    \]

    \item Third term:
    \[
    \mathcal{F} \left[ (|W|^2 - |V|^2) a(t) \right] = (|W|^2 - |V|^2) a(\omega)
    \]

    \item Right-hand side:
    \[
    \mathcal{F}[\Gamma(t)] = \Gamma(\omega)
    \]
\end{itemize}
\noindent
Substituting all the transformed terms into Eq.~(70):

\begin{align}
-\omega^2 a(\omega) + 2i\omega\Re(W) a(\omega) + \left( |W|^2 - |V|^2 \right) a(\omega) &= \Gamma(\omega)
\end{align}
\noindent
Factoring out $a(\omega)$:

\begin{equation}
\left[ -\omega^2 + 2i\omega\Re(W) + (|W|^2 - |V|^2) \right] a(\omega) = \Gamma(\omega)
\end{equation}

\subsection*{Final Expression for $a(\omega)$}

Solve for $a(\omega)$:

\begin{equation}
a(\omega) = \frac{\Gamma(\omega)}{-\omega^2 + 2i\omega\Re(W) + |W|^2 - |V|^2} 
\end{equation}
\noindent
The denominator is a quadratic in $\omega$, and its structure resembles a Lorentzian resonance, with:

\begin{itemize}
    \item Damping determined by $2\,\text{Re}(W)$,
    \item Resonant frequency shift influenced by $|W|^2$,
    \item Parametric interaction encoded in $|V|^2$.
\end{itemize}
\noindent
Peaks in the magnitude $|a(\omega)|$ occur near the frequencies where the denominator approaches zero — indicating resonance behavior.

\section*{Expression of $a(\omega)$ in Terms of Characteristic Roots}
\noindent
Starting from the frequency-domain solution:
\begin{equation}
a(\omega) = \frac{\Gamma(\omega)}{-\omega^2 + 2i\omega\,\text{Re}(W) + |W|^2 - |V|^2}
\end{equation}
\noindent
The aim is to factorize the quadratic denominator in terms of its characteristic roots.

\subsection*{Characteristic Equation}
\noindent
Defining the characteristic polynomial:
\begin{equation}
\lambda^2 - 2\,\text{Re}(W)\lambda + (|W|^2 - |V|^2) = 0
\end{equation}
\noindent
$\lambda_0$ and $\lambda_1$ are the roots. Then:
\[
\lambda_0 + \lambda_1 = 2\,\text{Re}(W), \quad \lambda_0 \lambda_1 = |W|^2 - |V|^2
\]

\subsection*{Factoring the Denominator}

\begin{align}
&-\omega^2 + 2i\omega\,\text{Re}(W) + |W|^2 - |V|^2 \\
&= (-i\omega)^2 + 2(-i\omega)\,\text{Re}(W) + |W|^2 - |V|^2 \\
&= (-i\omega + \lambda_0)(-i\omega + \lambda_1)
\end{align}

\subsection*{Final Expression}

Thus, rewriting $a(\omega)$ as:
\begin{equation}
a(\omega) = \frac{\Gamma(\omega)}{(-i\omega + \lambda_0)(-i\omega + \lambda_1)}
\end{equation}
\noindent
This form highlights the system's frequency response in terms of its complex poles $\lambda_0$ and $\lambda_1$, which govern resonance and damping behavior.

\section*{Derivation of \(\Gamma(t)\) and \(\Gamma(\omega)\) for the Heisenberg-Langevin equation}
\noindent
We know

\begin{equation}
\boxed{
\Gamma(t) = \frac{d F}{dt} + W^* F(t) - V F^\dagger(t).
}
\label{eq:Gamma_t}
\end{equation}
\noindent
Defining the Fourier transform:

\[
F(t) = \int_{-\infty}^\infty \frac{d\omega}{2\pi} e^{-i \omega t} F(\omega).
\]
\noindent
Then,

\[
\frac{d F}{dt} \rightarrow -i \omega F(\omega),
\quad
F^\dagger(t) \rightarrow F^\dagger(-\omega).
\]
\noindent
Hence, Eq.~\eqref{eq:Gamma_t} becomes

\begin{equation}
\boxed{
\Gamma(\omega) = (-i \omega + W^*) F(\omega) - V F^\dagger(-\omega).
}
\label{eq:Gamma_omega}
\end{equation}
\noindent
\\
Derivation of \(\Gamma(\omega)\) with three input ports:
\noindent
Given the time-domain operators and their Fourier transforms,
\begin{align}
a(t) &= \frac{1}{\sqrt{2\pi}} \int_{-\infty}^\infty d\omega\, a(\omega) e^{-i \omega t}, \\
c_j(t) &= \frac{1}{\sqrt{2\pi}} \int_{-\infty}^\infty d\omega\, c_j(\omega) e^{-i \omega t}, \quad j=1,2,3. 
\end{align}
\noindent
The operator \(F(t)\), representing environmental noise/input, can be decomposed into contributions from three ports:
\begin{equation}
F(t) = \sum_{j=1}^3 f_j(t),
\end{equation}
where each component is
\begin{equation}
f_j(t) = \sqrt{\gamma_j} e^{i \phi_j} c_{\mathrm{in}, j}(t).
\end{equation}
Here,
\begin{itemize}
    \item \(\gamma_j\) is the coupling rate for port \(j\),
    \item \(\phi_j\) is the associated phase,
    \item \(c_{\mathrm{in}, j}(t)\) is the input annihilation operator for port \(j\).
\end{itemize}

\medskip
\noindent
 Since \(c_3\) is driven by a pump at frequency \(\omega_p\), its operators appear with shifted frequencies \(\omega_p \pm \omega\) and an additional phase \(\phi_B\) due to the pump. Thus,
\begin{equation}
F(t) = 
\sqrt{\gamma_1} e^{i \phi_1} c_{\mathrm{in},1}(t)
+ \sqrt{\gamma_2} e^{i \phi_2} c_{\mathrm{in},2}(t)
+ \sqrt{\gamma_3} e^{i(\phi_B + \phi_3)} c_{\mathrm{in},3}(t).
\notag
\end{equation}
\noindent
Taking the Fourier transform,
\begin{equation}
\begin{aligned}
F(\omega) =\ 
& \sqrt{\gamma_1} e^{i \phi_1} c_{\mathrm{in},1}(\omega) \\
& + \sqrt{\gamma_2} e^{i \phi_2} c_{\mathrm{in},2}(\omega) \\
& + \sqrt{\gamma_3} e^{i(\phi_B + \phi_3)} c_{\mathrm{in},3}(\omega).
\end{aligned}
\end{equation}
\noindent
For the third port, due to the pump frequency, the relevant operators are :
\begin{equation}
c_{\mathrm{in},3}(\omega_p + \omega), \quad c_{\mathrm{in},3}^\dagger(\omega_p - \omega)
\end{equation}
Thus\\
\noindent
\begin{align}
F^\dagger(-\omega) 
&= \sqrt{\gamma_1} e^{-i \phi_1} c_{\mathrm{in},1}^\dagger(-\omega) 
+ \sqrt{\gamma_2} e^{-i \phi_2} c_{\mathrm{in},2}^\dagger(-\omega) \notag \\
&\quad + \sqrt{\gamma_3} e^{-i(\phi_B + \phi_3)} c_{\mathrm{in},3}^\dagger(\omega_p - \omega)
\end{align}
\\
\noindent
Starting from
\begin{equation}
\Gamma(\omega) = (-i \omega + W^*) F(\omega) - V F^\dagger(-\omega),
\end{equation}
\\
substituting the expressions for \(F(\omega)\) and \(F^\dagger(-\omega)\):
\begin{align}
\Gamma(\omega) =\ 
& (-i \omega + W^*) \big[ 
\sqrt{\gamma_1} e^{i \phi_1} c_{\mathrm{in},1}(\omega) 
+ \sqrt{\gamma_2} e^{i \phi_2} c_{\mathrm{in},2}(\omega) \nonumber \\
& \qquad\quad + \sqrt{\gamma_3} e^{i(\phi_B + \phi_3)} c_{\mathrm{in},3}(\omega_p + \omega) 
\big] \nonumber \\
& - V \big[ 
\sqrt{\gamma_1} e^{-i \phi_1} c_{\mathrm{in},1}^\dagger(-\omega) 
+ \sqrt{\gamma_2} e^{-i \phi_2} c_{\mathrm{in},2}^\dagger(-\omega) \nonumber \\
& \qquad\quad + \sqrt{\gamma_3} e^{-i(\phi_B + \phi_3)} c_{\mathrm{in},3}^\dagger(\omega_p - \omega) 
\big]
\end{align}
\noindent
Pulling out the common factors to match standard input-output theory conventions:
\begin{equation}
\boxed{
\begin{aligned}
\Gamma(\omega) =\ 
& -i \sqrt{2 \gamma_1} \big[ 
(-i \omega + W^*) e^{i \phi_1} c_{\mathrm{in},1}(\omega) 
\big] \\
& \quad + i \sqrt{2 \gamma_1} V e^{-i \phi_1} c_{\mathrm{in},1}^\dagger(-\omega) \\
& -i \sqrt{2 \gamma_2} \big[ 
(-i \omega + W^*) e^{i \phi_2} c_{\mathrm{in},2}(\omega) 
\big] \\
& \quad + i \sqrt{2 \gamma_2} V e^{-i \phi_2} c_{\mathrm{in},2}^\dagger(-\omega) \\
& -i 2 \sqrt{\gamma_3 B} \big[ 
(-i \omega + W^*) e^{i (\phi_B + \phi_3)} c_{\mathrm{in},3}(\omega_p + \omega) 
\big] \\
& \quad + i 2 \sqrt{\gamma_3 B} V e^{-i (\phi_B + \phi_3)} 
c_{\mathrm{in},3}^\dagger(\omega_p - \omega)
\label{112}
\end{aligned}
}
\end{equation}
\noindent
\begin{itemize}
    \item The factor \(2 \sqrt{\gamma_3 B}\) for the third port reflects the parametric drive strength \(B\).
    \item The phases \(\phi_j\) and \(\phi_B\) arise from the physical setup.
    \item The frequency shifts \(\omega_p \pm \omega\) on the third port reflect pump-induced sidebands.
\end{itemize}
\noindent
The output field can be expressed as
\begin{equation}
c_{\text{out}1}(\omega) = c_{\text{in}1}(\omega) - i\sqrt{2\gamma_1} e^{-i\phi_1} a(\omega),
\end{equation}
\\
One obtains:
\begin{align*}
c_{\text{out}1}(\omega) &= c_{\text{in}1}(\omega) - i \sqrt{2\gamma_1} e^{-i\phi_1} \cdot \frac{\Gamma(\omega)}{(-i\omega + \lambda_0)(-i\omega + \lambda_1)}
\end{align*}
\noindent
Defining the denominator:
\begin{align}
D(\omega) = (-i\omega + \lambda_0)(-i\omega + \lambda_1)
\end{align}
\noindent
Then:
\begin{align*}
c_{\text{out}1}(\omega) = \frac{D(\omega) c_{\text{in}1}(\omega) - i \sqrt{2\gamma_1} e^{-i\phi_1} \Gamma(\omega)}{D(\omega)}
\end{align*}
\noindent
Now plugging in $\Gamma(\omega)$ from Eq.~\eqref{112} and expanding the following expression:
\[
-i \sqrt{2\gamma_1} e^{-i\phi_1} \Gamma(\omega)
\]
\noindent
Breaking it into parts:\\
\\
\noindent
\textbf{Term 1:}
\begin{align*}
&- -i \sqrt{2\gamma_1} e^{-i\phi_1} \cdot \left(-i \sqrt{2\gamma_1} \left[ (-i\omega + W^*) e^{i\phi_1} c_{\text{in}1}(\omega) - V e^{-i\phi_1} c_{\text{in}1}^\dagger(-\omega) \right] \right) \\
&= 2\gamma_1 \left[ (-i\omega + W^*) c_{\text{in}1}(\omega) - V e^{-2i\phi_1} c_{\text{in}1}^\dagger(-\omega) \right]
\end{align*}
\noindent
\textbf{Term 2:}
\begin{align*}
&- -i \sqrt{2\gamma_2} e^{-i\phi_1} \cdot \left(-i \sqrt{2\gamma_2} \left[ (-i\omega + W^*) e^{i\phi_2} c_{\text{in}2}(\omega) - V e^{-i\phi_2} c_{\text{in}2}^\dagger(-\omega) \right] \right) \\
&= 2\gamma_1 \gamma_2 \left[ (-i\omega + W^*) e^{-i(\phi_1 - \phi_2)} c_{\text{in}2}(\omega) - V e^{-i(\phi_1 + \phi_2)} c_{\text{in}2}^\dagger(-\omega) \right]
\end{align*}
\noindent
\textbf{Term 3:}
\begin{align*}
& -i \sqrt{2\gamma_3} e^{-i\phi_1} \cdot 
\Bigg( -i \sqrt{2\gamma_3} B 
\Big[ (-i\omega + W^*) e^{i(\phi_B + \phi_3)} c_{\text{in}3}(\omega_p + \omega) \\
& \hspace{3.5cm} - V e^{-i(\phi_B + \phi_3)} c_{\text{in}3}^\dagger(\omega_p - \omega) 
\Big] \Bigg) \\
&= 2\gamma_1 \gamma_3 B 
\Big[ (-i\omega + W^*) e^{-i(\phi_1 - \phi_B - \phi_3)} c_{\text{in}3}(\omega_p + \omega) \\
& \hspace{3.8cm} - V e^{-i(\phi_1 + \phi_3 + \phi_B)} c_{\text{in}3}^\dagger(\omega_p - \omega) 
\Big]
\end{align*}
\\
\noindent
One obtains:
\begin{align}
c_{\text{out}1}(\omega) =\ 
& \frac{1}{(-i\omega + \lambda_0)(-i\omega + \lambda_1)} \bigg[ \nonumber \\
& \quad (-i\omega + \lambda_0)(-i\omega + \lambda_1) c_{\text{in}1}(\omega) \nonumber \\
& \quad - 2\gamma_1(-i\omega + W^*) c_{\text{in}1}(\omega) \nonumber \\
& \quad + 2\gamma_1 V e^{-2i\phi_1} c_{\text{in}1}^\dagger(-\omega) \nonumber \\
& \quad - 2\sqrt{\gamma_1 \gamma_2}(-i\omega + W^*) 
           e^{-i(\phi_1 - \phi_2)} c_{\text{in}2}(\omega) \nonumber \\
& \quad + 2\sqrt{\gamma_1 \gamma_2} V 
           e^{-i(\phi_1 + \phi_2)} c_{\text{in}2}^\dagger(-\omega) \nonumber \\
& \quad - 2\sqrt{2\gamma_1 \gamma_3} B (-i\omega + W^*) 
           e^{-i(\phi_1 - \phi_B - \phi_3)} 
           c_{\text{in}3}(\omega_p + \omega) \nonumber \\
& \quad + 2\sqrt{2\gamma_1 \gamma_3} B V 
           e^{-i(\phi_1 + \phi_3 + \phi_B)} 
           c_{\text{in}3}^\dagger(\omega_p - \omega) 
\bigg]
\label{114}
\end{align}

\noindent
Two important scenarios that are calculated in the next section are:
\begin{enumerate}
 \item Intermodulation gain: Only the input signal \( c_{\text{in}1}^\dagger(-\omega) \) is present. This corresponds to a classical signal at frequency \( \omega_p - \omega \), which leads to an output signal at frequency \( \omega_p + \omega \).
 \item Parametric gain: Only the input signal \( c_{\text{in}1}(\omega) \) is present. \\
 \noindent
This corresponds to a classical signal at frequency \( \omega_p + \omega \).
\end{enumerate}
\section{Intermodulation Gain}
\noindent
\\
Using the third term in Eq.~\eqref{114}, the Intermodulation Gain can be derived as :
\begin{equation}
G_I \equiv \frac{\left| c_{\text{out}, 1}(\omega) \right|^2}{\left| c_{\text{in}, 1}(-\omega) \right|^2} 
= \frac{4\gamma^2 \left| V \right|^2}{(\omega^2 + \lambda_0^2)(\omega^2 + \lambda_1^2)}
\end{equation}
The plot for intermodulation gain ($\omega=0$) is depicted in figure 3:
\begin{figure}[H]
   
    \begin{minipage}[b]{0.5\textwidth}
       
        \includegraphics[width=\textwidth]{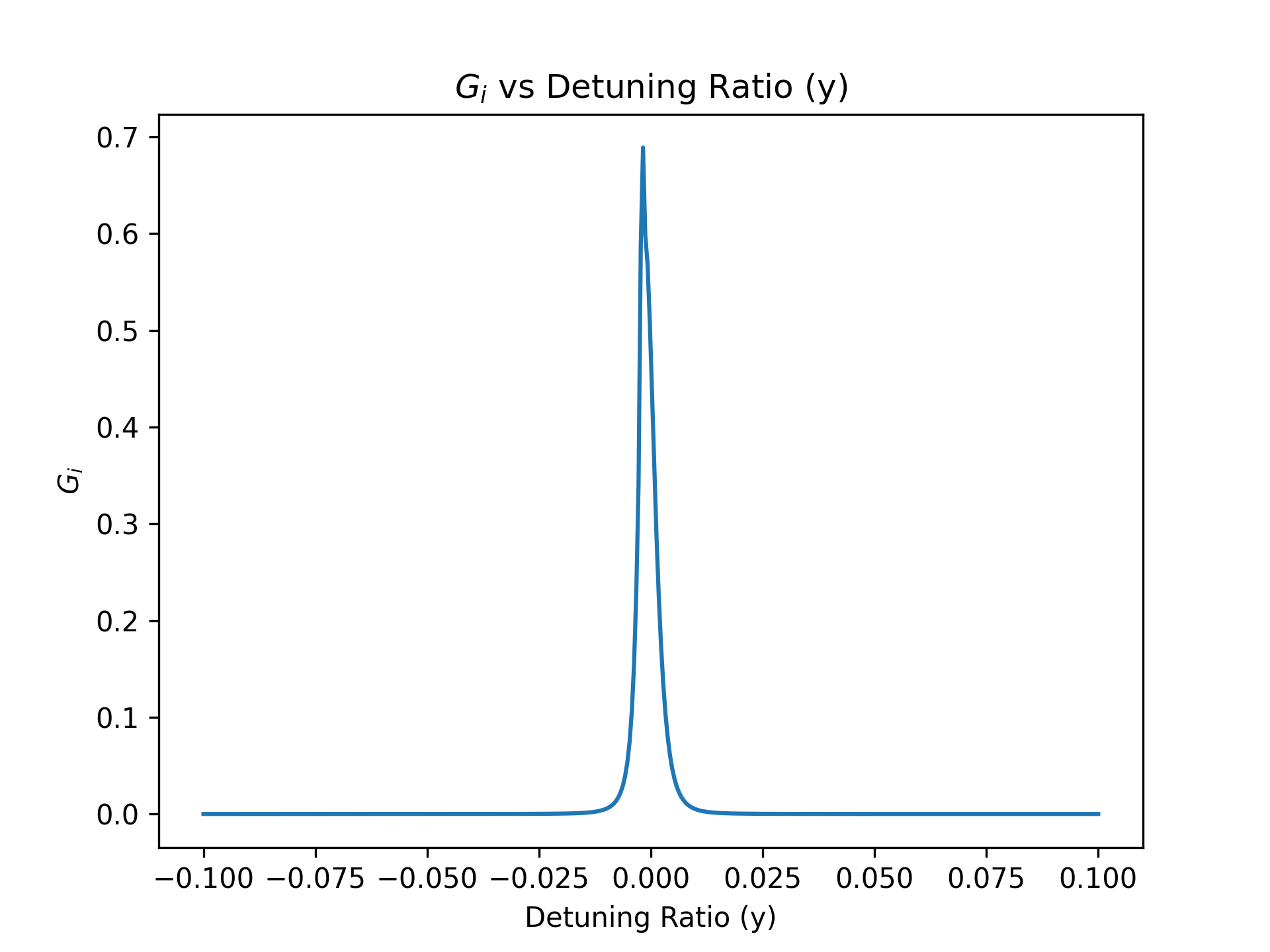}
        \label{fig:figure3}
    \end{minipage}
    \begin{minipage}[b]{0.5\textwidth}
       
        \includegraphics[width=\textwidth]{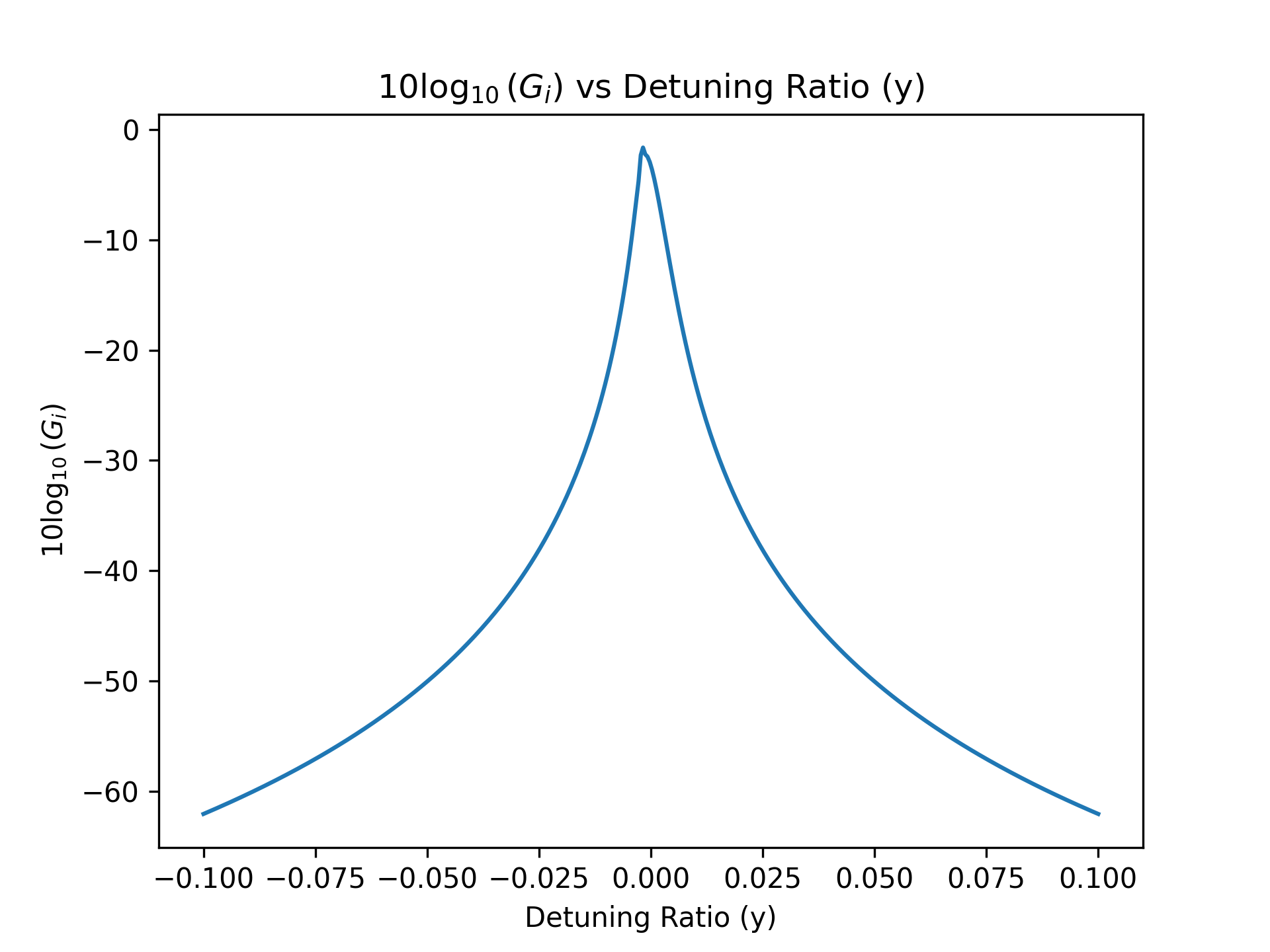}
        \label{fig:figure4}
    \end{minipage}
    \caption{Intermodulation Gain vs Detuning Ratio}
\end{figure}

\section{Parametric Gain}
\noindent
\\
The Parametric Gain can be derived as:\\
\begin{equation}
G_S \equiv \frac{\left| c_{\text{out}, 1}(\omega) \right|^2}{\left| c_{\text{in}, 1}(\omega) \right|^2} 
= \frac{\left| (-i\omega + \lambda_0)(-i\omega + \lambda_1) - 2\gamma(-i\omega + W^*) \right|^2}{(\omega^2 + \lambda_0^2)(\omega^2 + \lambda_1^2)}
\end{equation}
\\
For a pump amplitude of $b_{in,1}$= $0.8$$b_{in,1c}$, the plot for the parametric gain is depicted in Figure 4.
\begin{figure}[H]
    \begin{minipage}[b]{0.5\textwidth}
        \includegraphics[width=\textwidth]{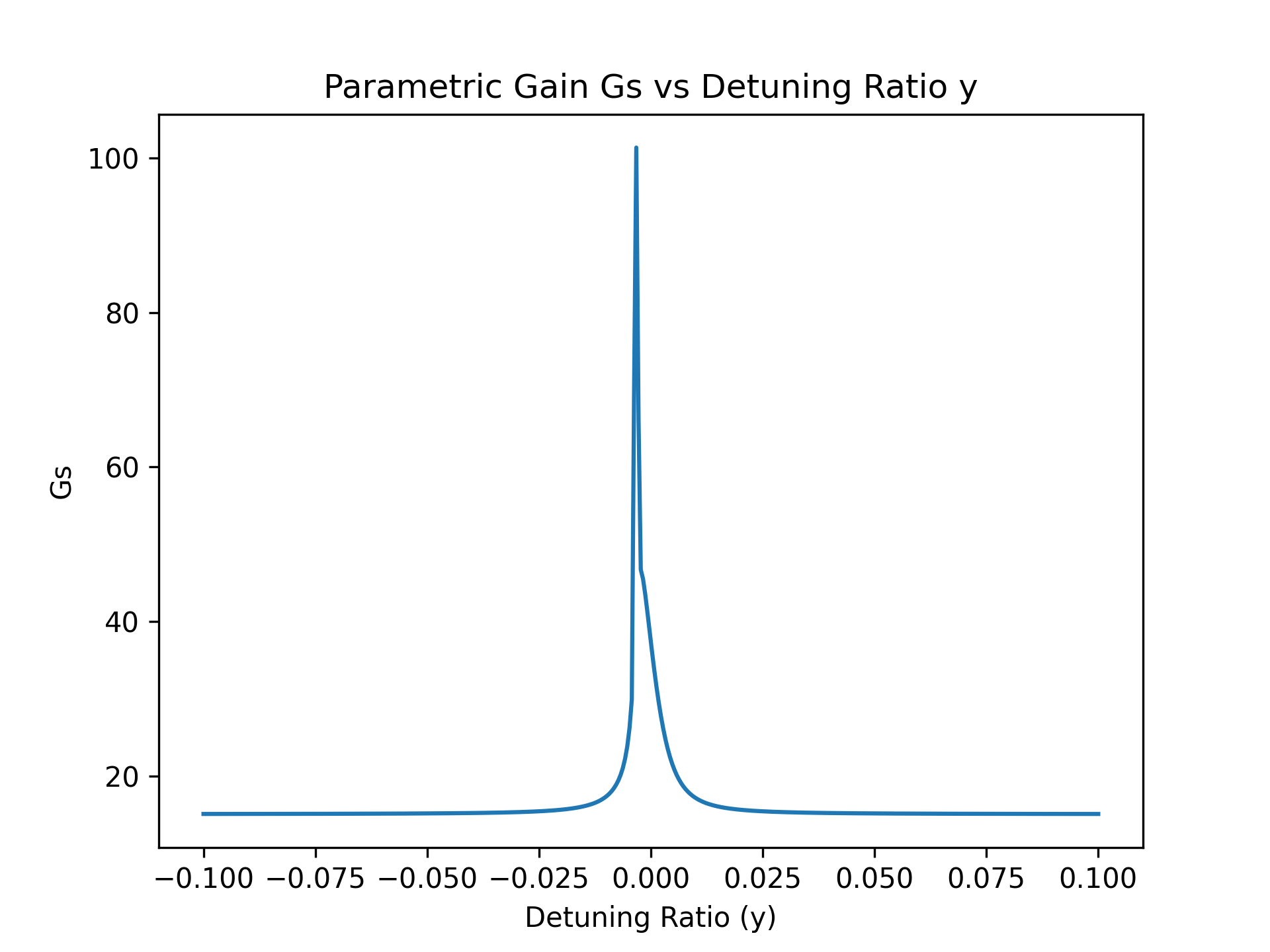}
        \label{Gain}
    \end{minipage}
    \begin{minipage}[b]{0.5\textwidth}
       
        \includegraphics[width=\textwidth]{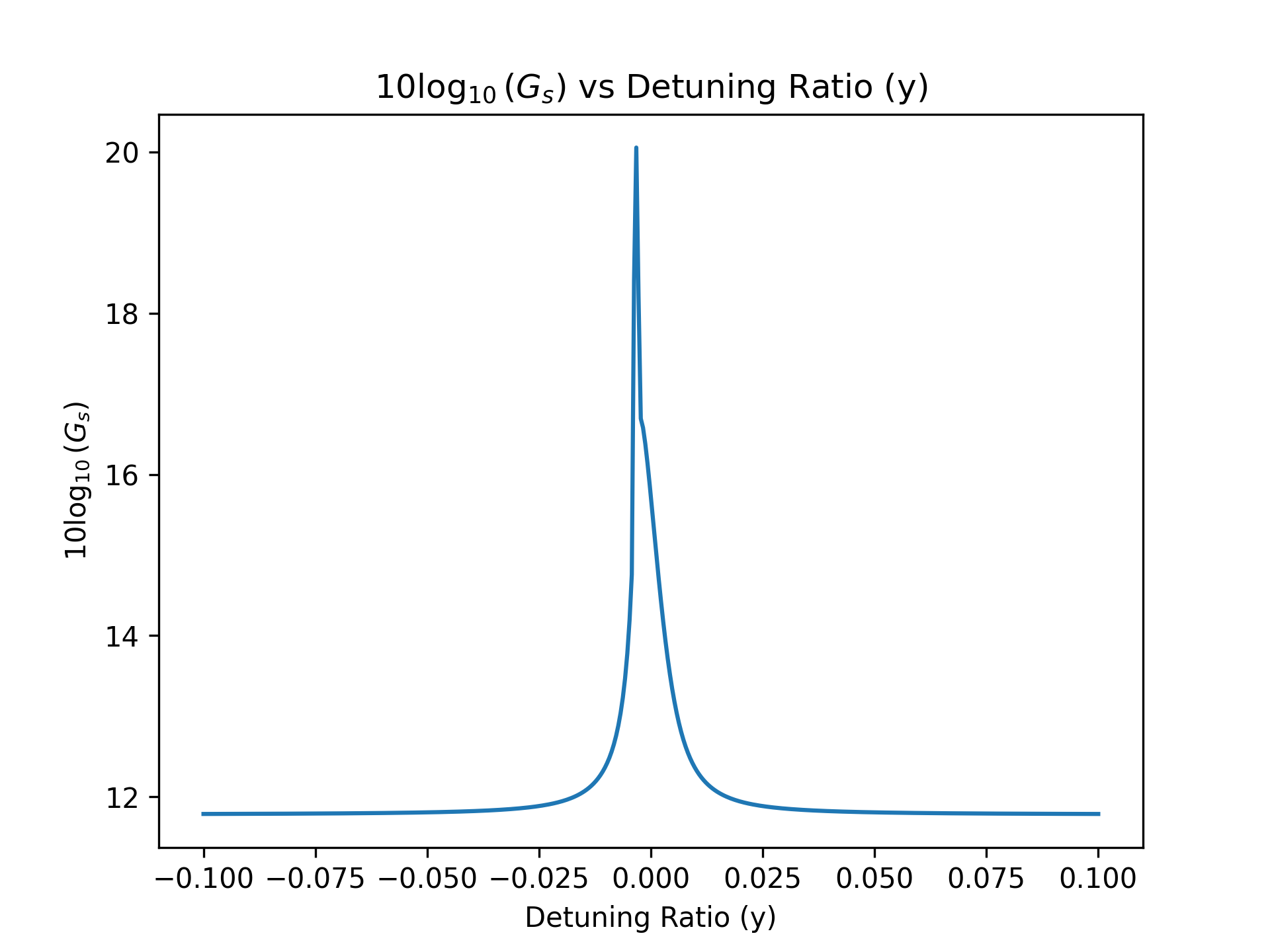}
    \end{minipage}
    \caption{Parametric Gain vs Detuning Ratio}
\end{figure}
\section{Conclusion}
The study presents a theoretical review of a cavity parametric amplifier that exhibits both Kerr nonlinearity and two-photon loss. The presence of a Kerr nonlinearity introduces a nonlinear dependence of the cavity resonance frequency on the intracavity field amplitude, while the two-photon loss adds a nonlinear damping mechanism. Expressions were derived in detail and visualized for several key quantities: the intracavity pump amplitude, which describes the field strength inside the resonator; the parametric gain, which quantifies amplification at the input signal frequency; and the intermodulation gain, which accounts for signal generation at frequencies different from the input due to nonlinear mixing effects. It is assumed that the pump saturation can be neglected. The analytical expressions serve as practical tools for fitting experimental data and extracting key model parameters, such as the Kerr coefficient \( K \), linear loss \( \gamma \), and two-photon loss rate \( \gamma_3 \). Importantly, the presence of two-photon loss significantly modifies the system's behavior. It increases the input power required to reach the bistable regime. The bistable regime ceases to exist when the two-photon loss rate exceeds a critical value, and the resonance curve no longer exhibits the characteristic S-shape, and the system becomes monostable. The discussion demonstrates that careful tuning of the Kerr nonlinearity and the two-photon loss rate is necessary for accessing and exploiting the nonlinear gain and bistability features in parametric amplifiers.


\begin{thebibliography}{9}
\bibitem{Yurke2005}
Yurke, Bernard, and Eyal Buks.  
\textit{Performance of Cavity-Parametric Amplifiers, Employing Kerr Nonlinearities, in the Presence of Two-Photon Loss}.  
ArXiv, 2005. \href{https://doi.org/10.48550/arXiv.quant-ph/0505018}{doi:10.48550/arXiv.quant-ph/0505018}.

\bibitem{Esposito2019}
Esposito, Martina, et al.  
\textit{Development and Characterization of a Flux-Pumped Lumped Element Josephson Parametric Amplifier}.  
EPJ Web of Conferences, vol. 198, 2019. \href{https://doi.org/10.1051/epjconf/201919800008}{doi:10.1051/epjconf/201919800008}.

\bibitem{Eichler2014}
Eichler, Christopher, and Andreas Wallraff.  
\textit{Controlling the Dynamic Range of a Josephson Parametric Amplifier}.  
EPJ Quantum Technology, vol. 1, no. 1, 29 Jan. 2014.  
\href{https://doi.org/10.1140/epjqt2}{doi:10.1140/epjqt2}. 

\bibitem{Castellanos2007}
Castellanos-Beltran, M. A., \& Lehnert, K. W.  
\textit{Widely tunable parametric amplifier based on a superconducting quantum interference device array resonator}.  
Applied Physics Letters, 91(8), 083509, 2007.  
\href{https://doi.org/10.1063/1.2773988}{doi:10.1063/1.2773988}.

\bibitem{Jeffrey2014}
Jeffrey, Evan, et al.  
\textit{Fast accurate state measurement with superconducting qubits}.  
Physical Review Letters, 112(19), 190504, 2014.  
\href{https://doi.org/10.1103/PhysRevLett.112.190504}{doi:10.1103/PhysRevLett.112.190504}. 

\bibitem{Walter2017}
Walter, Thomas, et al.  
\textit{Rapid high-fidelity single-shot dispersive readout of superconducting qubits}.  
Physical Review Applied, 7(5), 054020, 2017.  
\href{https://doi.org/10.1103/PhysRevApplied.7.054020}{doi:10.1103/PhysRevApplied.7.054020}. 

\bibitem{Macklin2015}
Macklin, Christopher, et al.  
\textit{A near–quantum-limited Josephson traveling-wave parametric amplifier}.  
Science, 350(6258), 307–310, 2015.  
\href{https://doi.org/10.1126/science.aaa8525}{doi:10.1126/science.aaa8525}.  

\end{thebibliography}
\end{document}